\documentclass[preprint]{aastex}

\usepackage{ccaption}
\usepackage{epsfig}
\usepackage{amsmath}
\usepackage{multirow}

\shorttitle{A Bayesian approach to comparing theoretic models to observational data: A case study from solar flare physics}
\shortauthors{Adamakis et al.}

\begin{document}

\title{A Bayesian approach to comparing theoretic models to observational data: \\ 
A case study from solar flare physics}

\author{S. Adamakis\altaffilmark{1}}

\author{C.~L. Raftery\altaffilmark{3}}

\author{R.~W. Walsh\altaffilmark{2}}
\altaffiltext{1}{Decision Science, Lloyds Banking Group, 155 Bishopsgate, London, EC2M~3TQ, UK}
\altaffiltext{2}{Jeremiah Horrocks Institute for Astrophysics and Supercomputing, University of Central Lancashire, Preston, PR1~2HE, UK}
\author{P.~T. Gallagher\altaffilmark{3}}
\altaffiltext{3}{Astrophysics Research Group, School of Physics, Trinity College, Dublin, Dublin 2, Ireland}

\begin{abstract}
Solar flares are large-scale releases of energy in the solar atmosphere, which are characterised by rapid changes in the hydrodynamic properties of plasma from the photosphere to the corona. Solar physicists have typically attempted to understand these complex events using a combination of theoretical models and observational data. From a statistical perspective, there are many challenges associated with making accurate and statistically significant comparisons between theory and observations, due primarily to the large number of free parameters associated with physical models. This class of ill-posed statistical problem is ideally suited to Bayesian methods. In this paper, the solar flare studied by \citet{raftery08} is reanalysed using a Bayesian framework. This enables us to study the evolution of the flare's temperature, emission measure and energy loss in a statistically self-consistent manner. The Bayesian-based model selection techniques imply that no decision can be made regarding which of the conductive or non-thermal beam heating play the most important role in heating the flare plasma during the impulsive phase of this event.
\end{abstract}


\section{Introduction}
\label{introduction}

Solar flares are triggered by the instability of the magnetic field. This can result in the direct or indirect heating of chromospheric plasma leading to the process known as ``chromospheric evaporation'' \citep{milligan06a, milligan06b}. There are two ways to provoke evaporation (chromospheric plasma upflow): using thermal energy and/or using non-thermal energy according to the energy release mechanisms \citep{mariska93}. In the thermal energy model, heat is unleashed in the coronal portion of the loop (possibly because of magnetic reconnection). Then the thermal energy is carried downward to the upper chromosphere via conduction, where the deposited energy heats the plasma and stimulates it to move slowly upward. In the non-thermal energy model, the energy release (again in the corona) is in the form of a non-thermal electron beam. Electrons then collide with dense chromospheric plasma, which they heat, causing the plasma to expand both upward at very high speeds and downward at lower speeds.

This paper will focus on the statistical analysis of the observations outlined in \citet{raftery08} and will be compared with results from the Enthalpy Based Thermal Evolution of Loops (EBTEL) model \citep{kpc08}. Comparing to \citet{raftery08} this research will handle the statistics with care to derive which of the thermal or non-thermal heat flux is more dominant, during a single flare event, using different model comparison techniques \citep{kass93,kgr89,kr95,rafb96,clyde07,aitkin09}. Furthermore, the method used here treats the uncertainties for the time constraints with equal respect as the uncertainties of the temperature and emission measure, which are difficult to incorporate into our analysis regarding Classical statistics. The thermal heat flux is proportional to the non-thermal heat flux plus a constant background heat flux. The parameter set is extended with the addition of the radius of the loop in order to make the model more realistic. Last but not least, the ratios between the radiative loss rate of the transition and the corona, the average coronal and apex temperature, and the coronal base and apex temperature are assumed to be free parameters so that their values will be decided by the data. For more information about these parameters refer to Section \ref{model_parameters}.

One advantage of the EBTEL model is that we can incorporate both a direct and a non-thermal heat input, which is not restricted by the data. Thus, in order to produce a temperature, density and/or emission measure profile, a specific form for the non-thermal heat flux function must be assumed. There appears to be a connection between the heating function form and the temperature profile, more so than with the emission measure profile. For example, a sudden increase in the heat input will lead to an abrupt uplift in the temperature, whereas a flat heating function will lead to a smoother temperature profile. Hence, it is important to understand and evaluate the impact any changes in the thermal and non-thermal heat flux have on the plasma evolution. Several forms of thermal heat fluxes have been tested in \citet{adamakis09}. Here we present only two of them: Half Gaussian profile and Full Gaussian profile. The Half Gaussian profile represents a sudden switching off of an electron beam, whereas the Full Gaussian profile imitates a gradual one. The rest of the paper has been structured as follows: Section \ref{previous_work} comments on the \citet{raftery08} observations and some assumptions in the EBTEL model. Section \ref{model_parameters} discusses the details of the analysis regarding the data distribution and the parameter set. Section \ref{priors} addresses to the choice of the parameters in the prior distributions. Section \ref{sec:MCM} briefly presents the model selection methods applied here. Section \ref{results_raftery} presents results of different models and parameter estimations. Finally, Section \ref{discussion} summarise the findings from this study and presents how this work can be further progressed.

\section{Previous Work}
\label{previous_work}

\subsection{Observations of a C-Class Solar Flare}

\begin{figure}
   \centering
   \includegraphics[scale=0.44]{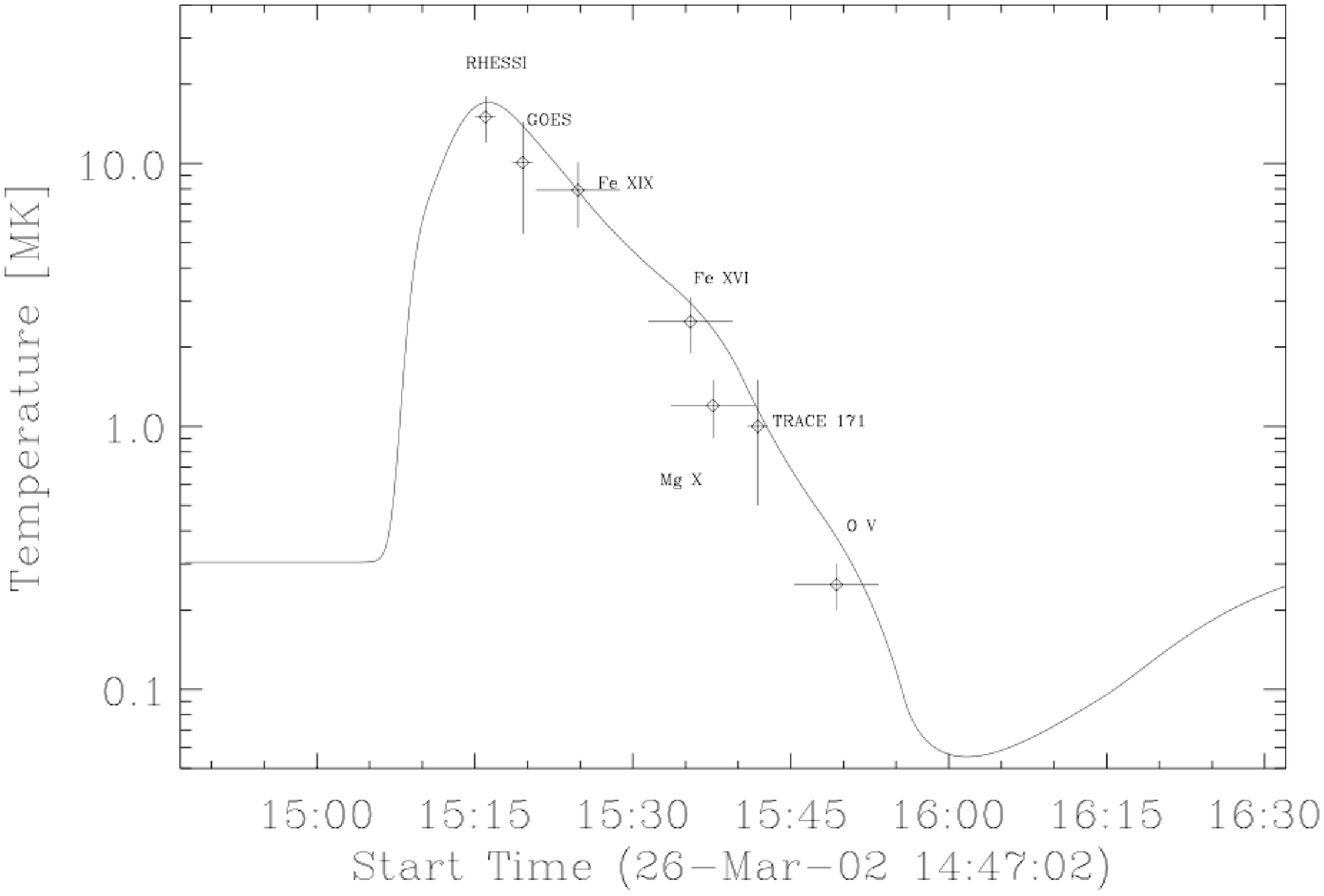} \\
   \includegraphics[scale=0.44]{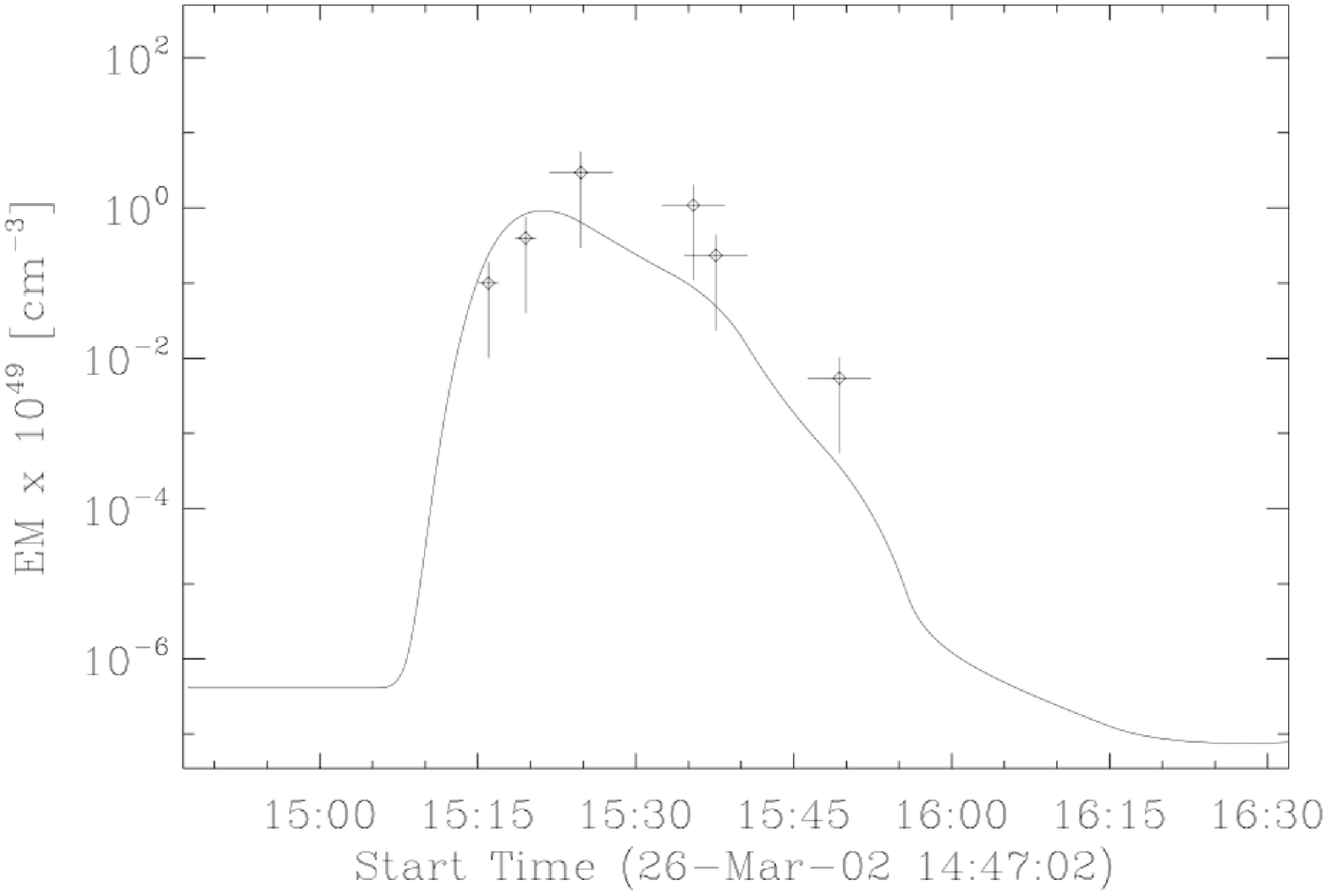} \\ 
   \caption{Temperature and emission measure observations of a C3.0 solar flare reported in \citet{raftery08}. \emph{Top}: The EBTEL temperature evolution that best reproduced the observations, according to \cite{raftery08}. Different instruments were employed to measure the temperature at different times. \emph{Bottom}: The EBTEL model and observed emission measure evolution. TRACE observations were not included.}
   \label{raftery_EM_T_t}
\end{figure}

The temporal evolution of temperature and emission measure in a C3.0 solar flare observed on March 26, 2002 has been analysed. During a typical solar flare, the temperature rises from less than one MK up to $\sim 20$ MK, and then cools down to the pre-flare temperatures. In the example under consideration here, \citet{raftery08} used different instruments to track this evolution: the Reuven Ramaty High Energy Solar Spectroscopic Imager (RHESSI; $>5$ MK), GOES-12 ($5-30~$MK), the Transition Region and Coronal Explorer (TRACE $171$ \AA\ ; 1 MK), and the Coronal Diagnostic Spectrometer (CDS; $\sim 0.03-8$ MK). A notation should be made that TRACE data were not included in the emission measure analysis. This is as a result of the instrument being sensitive to multiple emission lines, making it complex to define the contribution function. The reader is referred to \citet{raftery08} which outlines the reasons why and at which particular time of the flare each particular instrument was employed. Figure \ref{raftery_EM_T_t} depicts a visual display of the observation results together with the underlying uncertainties.

The flare began with evidence of pre-flare heating at its onset. This was followed by explosive chromospheric evaporation during the impulsive phase and gentle chromospheric evaporation during the early decay phase. It is believed that the plasma reached a peak temperature of just more than $13$ MK in approximately 10 min. Conduction losses dominated over radiative losses for the initial $\sim~300$ s of the post-flare decay, whereas for the next $\sim 4000$  s, radiative losses prevailed. \citet{raftery08} concluded that approximately equal direct and non-thermal heating mechanisms produced the data observed according only to an estimation of the ratio between the thermal and non-thermal heating rate, which is of course very critical. The best fit and parameter intervals were derived using an acceptable fit to the data by eye.

\subsection{The EBTEL Model}

The Enthalpy-Based Thermal Evolution of Loops (EBTEL) model \citep{kpc08} is a zero-dimensional hydrodynamic (0D HD) model that takes into account the significance of introducing the enthalpy into the system. The difference with the 1D HD model is that the 1D HD model provides us with temperature, density, and velocity profiles along a given magnetic field line as time evolves (that is temperature, density, and velocity are a function of both space and time), whereas 0D HD models describe the average evolution of these values along a coronal strand as a function of time only. Subsequently, 0D HD models require less computing time than 1D models, but at the expense of losing the spatial resolution. In spite of this, the EBTEL code has been used in a wide range of studies regarding different heat input forms \citep[see][for more details]{kpc08}. The importance of 0D models compared to the 1D models is that they can give similar results for the plasma response to a sudden heat input, despite the fact that they use up to four orders of magnitude less computing time.

To reflect the physical processes taking place within the standard flare model, the EBTEL model allows for both thermal and non-thermal heating of the plasma in the system. 
It has been noted that proton precipitation is not included \citep[see][]{asch05}. Proton beams are expected to excite strong kinetic Alfv\'en waves. The turbulence caused by kinetic Alfv\'en waves contains enough energy to produce the non-thermal velocities observed in flares, and therefore could contribute to impulsive primary plasma heating between the reconnection regions and the flare footpoints. 

The governing equations of the EBTEL model for the coronal part of the loop are \citep{kpc08}:
\begin{eqnarray}
\label{PDE1}
%\frac{d\bar{P}}{dt} & \approx & \frac{2}{3} \left[ \bar{Q} - \bar{n}^2 \Lambda(\bar{T}) (1+c_1) \right], 
%\\
\frac{d\bar{P}}{dt} & \approx & \frac{2}{3} \left[ \bar{Q} - \bar{n}^2 \Lambda(\bar{T}) (1+c_1) - \frac{\mathcal{F}}{L} \left( 1 - \frac{3}{2} \frac{k \bar{T}}{\mathcal{E}} \right) \right], 
\\
%\frac{d\bar{n}}{dt} & = & -\frac{c_2}{5 c_3 k L \bar{T}} (F_0 + \mathcal{R}_{tr}), 
%\\
\label{PDE2}
\frac{d\bar{n}}{dt} & = & -\frac{c_2}{5 c_3 k L \bar{T}} (F_0 + \mathcal{R}_{tr}) + \frac{\mathcal{F}}{\mathcal{E} L} \left(1 - \frac{c_2}{5 c_3} \frac{\mathcal{E}}{k \bar{T}} \right), 
\\
\label{PDE3}
\frac{d\bar{T}}{dt} & \approx & \bar{T} \left(\frac{1}{\bar{P}} \frac{d\bar{P}}{dt} - \frac{1}{\bar{n}} \frac{d\bar{n}}{dt} \right) , 
\end{eqnarray}
where $P$ is the pressure, $t$ the time, $Q$ the direct (thermal) heating rate, $n$ the electron number density, $T$ the temperature, $\Lambda(\cdot)$ the optically thin radiative loss function, $\mathcal{F}$ the non-thermal energy flux ($\mathcal{F} = \mathcal{E} \mathcal{J}$, with $\mathcal{E}$ the mean energy of the accelerated (non-thermal) electrons impinging the chromosphere in keV and $\mathcal{J}$ the nonthermal particle flux), $k$ the Boltzmann's constant ($\approx~1.38~\times~10^{-16}$~erg~K$^{-1}$), $L$ the length of the loop from the coronal base to the apex, $F_0$ the thermal conduction (heat flux) at the beginning of the coronal region, $R_{tr}$ the radiative cooling rate at the transition region, $c_1=\frac{R_{tr}}{R_c}$ the ratio of the radiative loss function between the transition region and the corona, $c_2=\frac{\bar{T}}{T_a}$ the ratio between the average temperature at the coronal part of the loop and the temperature at the apex of the loop, $c_3=\frac{T_0}{T_a}$ the ratio between the temperature at the base of the corona and the temperature at the apex of the loop and the over bars indicate spatial averages along the coronal section of the loop. Furthermore, the transition region is treated separately. 

Deviations of the observations from the model could be caused by two distinct reasons: First, the EBTEL model treats the effects of the non-thermal electron beam in a very simple way. It is assumed that all the beam's energy goes into evaporating plasma upwards into the loop, which is reasonable for gentle evaporation. However, for explosive evaporation, some of the beam energy will be used to drive chromospheric downflows, making the total observed non-thermal energy larger than that predicted by the EBTEL model. Second, the flare loop is almost certainly constructed of many strands that are heated at different times. \citet{raftery08} assume the flaring loop consists of individual strands, all heated simultaneously. Therefore, they model the loop as one ``fat strand''. However, some strands are likely to be heated at a different time compared to this solid structure.

\section{Defining the Model Parameters}
\label{model_parameters}

\subsection{Data Distribution}
\label{data_distr}

The errors in temperature come from the width of the contribution function for the particular emission line (for the CDS points) and from the width of the instrument response function for GOES and TRACE. The RHESSI temperature error comes from the uncertainty in the fit to the Maxwell Boltzmann distribution. 

As there is no information about the form of the errors, it is assumed that the errors between the observed temperatures and emission measures and those predicted from the model are normally distributed with mean zero and standard deviation derived from the error bars. In this particular case study we assume ``$3 \sigma$'' belief, \emph{i.e} the distance between the mean of the distribution and the upper or lower error bar is three standard deviations of the normal distribution \citep[see][for a link between error bars and statistical distributions]{adamakis08}.
% \citep[we assume ``$3 \sigma$'' belief; see][]{adamakis08}. 
The reason for choosing a Normal distribution as the likelihood function is two-fold: (\emph{i}) the error bars are symmetric and intuitively a symmetric likelihood is needed, and, (\emph{ii}) it is the standard likelihood distribution that is applied in many parametric models in the absence of additional information. 

This means that the combined likelihood which contains both temperature and emission measure information is of the form: 
\begin{eqnarray*}
p(\mathbf{D} | \mathbf{P}) & = & \frac{1}{ \left( 2\pi \right)^{(n_1+n_2)/2} \left( \prod_{i=1}^{n_1} {\sigma_{i1}} \right) \left( \prod_{i=1}^{n_2} {\sigma_{i2}} \right) } \\ 
& & \exp{\left[ - \sum_{i=1}^{n_1} {\frac {\left(T_i - \widehat T_i(\mathbf{P}) \right)^2} {2\sigma_{i1}^2}} - \sum_{i=1}^{n_2} {\frac {\left(EM_i - \widehat {EM}_i(\mathbf{P}) \right)^2} {2\sigma_{i2}^2}}  \right]} \\ 
& & \mathcal{I}\left(\max{(T)}<3\times10^7\right), 
\end{eqnarray*}
where $\mathbf{D}=\left( \mathbf{T}, \mathbf{EM} \right)$ with $\mathbf{T}=(T_1, \ldots, T_{n_1})$ the temperature data-points and $\mathbf{EM}=(EM_1, \ldots, EM_{n_2})$ the emission measure data-points, $\widehat T_i$ and $\widehat {EM}_i$ are the temperatures and emission measures respectively proposed by the EBTEL model, $\mathbf{P}$ is the parameter set (see end of Section \ref{model_parameters}), $n_1$ is the number of observed temperature values, $n_2$ is the number of observed emission measure values, $\sigma_{i1}$ are the standard deviations of the temperature errors, $\sigma_{i2}$ are the standard deviations of the emission measure errors and $\mathcal{I}(\cdot)$ is the indicator function which is given by: 
\begin{displaymath}
\mathcal{I}\left(\max{(T)}<3\times10^7\right) = \left\{
\begin{array}{ll}
1, & \quad $~if~$ \max{(T)}<3\times10^7~\mbox{K} \\
0, & \quad $~otherwise,$
\end{array}
\right.
\end{displaymath}
where $\max{(T)}$ is the maximum temperature value of the cooling curve. The temperature profile is restricted to not exceed 30 MK because it is believed that temperatures above this threshold will create unphysical data. Temperature and electron density ($T$ and $n$ respectively) are calculated numerically from Equations (\ref{PDE1}) -- (\ref{PDE3}). First order finite difference methods are employed to solve numerically the partial differential equations, after providing initial values for temperature, density and pressure. Emission measure ($EM$) can be computed from $2 n^2 \pi r^2 L$, where $r$ is the dimensionless radius of the loop and $L$ is the dimensionless length of the loop.

\subsection{Parameters for Non-Thermal Heat Flux}
\label{parameters}

As discussed in Section \ref{introduction}, one of our aims is to compare which of the Half Gaussian and Full Gaussian functions for the non-thermal heat flux fits better to the data observed. In the case of the Half Gaussian the non-thermal heat flux function is given by the form: 
\begin{displaymath}
\mathcal{F}(t) = \left\{ 
\begin{array}{ll}
\frac{\mathcal{A}}{\sqrt{2\pi} \sigma_1} \exp{\left[-\frac{(t-\mu)^2}{2\sigma_1^2}\right]}, \quad t \le \mu \\
0, \quad t > \mu,
\end{array}
\right.
\end{displaymath}
where $t$ is the time, $\mu$ is the time that maximise the non-thermal heat flux function (which can be adopted as a parameter), $\sigma_1$ is a value that defines the width of this distribution (which can be adopted as a parameter as well) and $\mathcal{A}$ is the amplitude of the function (parameter). For the total non-thermal heat function $F_{tot}$ we have: 
\begin{displaymath}
\mathcal{F}_{tot} = \int_0^\infty \mathcal{F}(t)dt = \int_0^\mu \mathcal{F}(t)dt = P\left( 0 \le t \le \mu  \right) \mathcal{A},
\end{displaymath}
where $P(\cdot)$ defines the probability. We have $P(0<t<\mu)=p \le 0.5$ if and only if $\sigma_1 = \frac{\mu}{\Phi^{-1}\left[\frac{1+2p}{2} \right]}$, where $\Phi^{-1}(\cdot)$ is the cumulative distribution function of the Gaussian distribution with mean 0 and standard deviation 1. In the case that $\mu$ is big and $\sigma_1$ is small (\emph{e.g.} $\sigma_1 \le \mu /3$), then $P\left( 0 \le t \le \mu  \right)~\approx~\frac{1}{2}$. For these examples, we can assume $\mathcal{F}_{tot}=\frac{\mathcal{A}}{2}$. Moreover, if $\mathcal{F}'_{tot}$ is the dimensionless total non-thermal heat flux, then $\mathcal{F}_{tot}=\mathcal{F}'_{tot} \times 10^9$ ergs $\mbox{cm}^{-2}$.

In the case of the Full Gaussian the heat flux function is given by the form: 
\begin{displaymath}
\mathcal{F}(t) = \frac{\mathcal{A}}{\sqrt{2\pi} \sigma_1} \exp{\left[-\frac{(t-\mu)^2}{2\sigma_1^2}\right]}. 
\end{displaymath}
We have $P(t>0)=p$ (with $0.5 < p \le 1$) if and only if $\sigma_1 = \frac{\mu}{\Phi^{-1}(p)}$. If again, \emph{e.g.}, $\sigma_1 \le \mu/5$, then $P(t>0) \approx 1$ and for these cases we can assume $\mathcal{F}_{tot}=\mathcal{A}$.

\subsection{Parameters with the EBTEL Model}

The thermal heating rate is assumed to be of the form: 
\begin{equation}
\label{alpha_param_hest_rate}
Q(t) = \alpha \mathcal{F}_1(t) + B.  
\end{equation}
Here $Q$ is the direct heating rate; $B=B' \times 10^{-5}$ ergs $\mbox{cm}^{-3}$ $\mbox{s}^{-1}$ is a constant background heating rate that is occurring ($B'$ is dimensionless and is included as a parameter); $\mathcal{F}_1(t)$ is the non-thermal heating rate with $\mathcal{F}_1(t) = \frac{\mathcal{F}(t)}{L}$ where $L=L' \times 10^9$ cm is the loop's length from the top of the chromosphere to the apex (parameter); and $\alpha$ is a factor that defines which of the two heating functions (thermal or non-thermal) is dominant assuming that the background heating rate is negligible comparing the other two heating sources. Under this assumption, if $\alpha = 1$ we have equal amounts of thermal and non-thermal heating, if $\alpha>1$ then thermal heat flux is more dominant, whereas if $\alpha<1$ then non-thermal heat flux is more dominant. In particular, we will be examining comparisons of the form: $H_0: \alpha \ne 1$, $H_1: \alpha=1$, $H_2: \alpha>1$ and $H_3: \alpha<1$. Finally, we include the dimensionless radius of the loop $r'$ (assuming the loop to be homogeneous), with $r=r'~\times~10^9~$cm, as another parameter into our analysis. The importance of $r'$ is that we calculate $2 n^2 \pi r^2 L$ in order to compare it with the observed emission measure values. 

Initially, when this data-set was firstly assigned for analysis, it was suggested that $c_i$ should be fixed to the values proposed in \cite{kpc08}. However, it was interesting to leave them as parameters to check whether: \emph{i}) we will get data profiles with values close to those proposed in \cite{kpc08}, \emph{ii}) if not, whether these changes affect other parameters and subsequently model comparisons. Therefore, in Section \ref{c_i fixed}, these values are assumed to be fixed to given numbers, whereas in Section \ref{c_i parameters} they are assumed to be free parameters which can be determined by the data.

\subsection{Dealing with time}

The data values have uncertainties (error bars) not only on the $y$ axis (temperature and emission measure), but on the $x$ axis as well (time). These time errors were calculated as being the width of the spline interpolated light curves. There have been some attempts in the past to deal with problems of error bars in both axes \citep{wins46,berk50,jaynes99}. Reviews of errors in variable models can be found in \citet{cas90, cheng99, fuller87}. For extragalactic astronomy applications of these methods the reader is referred to \citet{ab96, kelly07, patriota09}. 

A Bayesian solution to the problem using the reversible jump MCMC algorithm can be found in \citet{hend00}. The latter application is utilised in the current paper and according to this, the \emph{true} time\footnote{The \emph{true} value of a variable ({\it e.g.} $T_{true}$) is based on the \emph{observed} value of that variable ($T_{obs}$) with the addition of an error ($\Delta T$), {\it i.e.} $T_{true} = T_{obs} + \Delta T$.} should be included as another parameter. 

$\\$

All in all, the parameter set would be: $\mathbf{P}=\left(\mathbf{P_1}, \mathbf{P_2} \right)$, with $\mathbf{P_1}=(t_1, \ldots, t_{n_1})$, where $t_i$ is the time for the $i$th observation, and $\mathbf{P_2} = (L', \mathcal{F}'_{tot}, \mu, \sigma_1, \alpha, B', r')$ for Section \ref{c_i fixed} and $\mathbf{P_2} = (L', \mathcal{F}'_{tot}, \mu, \sigma_1, \alpha, B', r', c_1, c_2, c_3)$ for Section \ref{c_i parameters}.

\section{Defining the Priors for the Parameters}
\label{priors}

A major difference between Classical statistics and Bayesian statistics is that the former does not consider any prior information for the parameters, whereas the latter can incorporate any available information about the parameters before we observe the data. Further discussion on choosing priors can be found in \citet{kw96}, while the importance of priors in model comparison can be found in \citet{kass93} and \citet{kgr89}. 

There are three main ways of choosing a prior: 
\begin{enumerate}
\item Subjective: the prior expresses the experimenter's personal probability that a parameter lies within a specified range. 
\item Objective and informative: the experimenter may have information or historical data prior to the experiment being undertaken. 
\item Non-informative: expresses ignorance about the value of a parameter and is usually dominated by the likelihood function. 
\end{enumerate}

For the current analysis, we incorporated subjective prior wherever it was possible and non-informative in any other case. It is worth noting that in case studies like this, prior distributions can prevent parameters from taking unphysical values or values that do not agree with what we observe. In this case, prior belief refers to any guess we have about a particular parameter that is related to the current observations under consideration, rather than prior belief from previous studies.  

The belief we have \emph{before} we undertake the analysis in Section \ref{results_raftery} is summarised in Table \ref{c_i prior}. We conclude from the observations that $L'$ should be around 3 and probably between 2 and 4. We have the option to give more weight to the value of 3 and less weight as we move away from that value. Hence, we assume that $L'$ has a Gamma distribution, $L' \sim \mathcal{G}(37.64, 0.08)$, where $\mathcal{G}(\alpha, \beta)$ is a Gamma distribution with shape parameter $\alpha$, scale parameter $\beta$ and mean $\alpha \beta$, which will give a $95\%$ probability between 2 and 4 with mode at 3. For the total non-thermal energy $\mathcal{F}'_{tot}$ we do not have any information at all, therefore $\pi \left(\mathcal{F}'_{tot} \right) \propto 1$, where $\pi(\cdot)$ denotes the prior probability density function.

\begin{table}
\begin{center}
\caption{Prior belief about the parameters.}
\label{c_i prior}
\begin{tabular}{c|ccc}     % define the column alignment
                           % l: left, c: center, r: right
  \hline\hline                   % horizontal line
  & mode & $95\%$ probability & prior distribution \\
  \hline\hline
$L'$       & 3 & [2,4] & $\mathcal{G}(37.64, 0.08)$ \\
$F'_{tot}$ & NA & NA & $\propto 1$ \\
$\mu$      & NA & NA & $\propto \mathcal{I}(\mu \le 884)$ \\
$\sigma_1$  & 100 & [20,200] & $\mathcal{G}(6.78, 17.28)$ \\
$\alpha$  & NA & NA & $\propto 1$ \\
$B'$       & NA & NA & $\propto 1$ \\
$r'$       & 0.49 & [0.20,1.24] & $\mathcal{G}(5.23,0.12)$ \\
$c_1$      & NA & NA & $\mathcal{G}(1.70, 2.36)$ \\
$c_2$      & NA & NA & $\mathcal{U}(0, 1)$ \\
$c_3$      & NA & NA & $\mathcal{U}(0, 1)$ \\
  \hline\hline
\end{tabular}
\end{center}
\end{table}

We assume that the flare observations are from the cooling phase of the temperature profile; from the highest temperature lines downwards, the temperature drops continually. Thus, it is very likely that the main flare energy release will occur at most up until the first observation. For this reason we choose the mean of the non-thermal heat flux ($\mu$) to have an upper limit at the lower error bar of the first observation (which is 884 s after the beginning of the observational period). Since we do not have any other belief about giving weights to any specific values we assume $\pi(\mu) \propto \mathcal{I}(\mu \le 884)$, where $\mathcal{I}(\cdot)$ is the indicator function (see Section \ref{data_distr}). Also, $\sigma_1 \sim \mathcal{G}(6.78, 17.28)$ will give $95\%$ probability for the one standard deviation of the heating function to be between 20 and 200 with mode at 100. Since we do not not have any knowledge about the background heating, we can assume an improper prior, \emph{e.g.} $\pi(B') \propto 1$. In Section \ref{c_i parameters} we use $\mathcal{G}(1.70, 2.36)$ as a prior for $c_1$ and the uniform prior $\mathcal{U}(0, 1)$ for $c_2, c_3$. The prior for $c_1$ will give $\sim 99.90\%$ probability for values below 20 and those for $c_2$ and $c_3$ come naturally as they will impose an upper limit of one, without favouring any values below unity. Note that the prior for $c_2$ can be conservative, as there might be evidence that this number is close to the value 0.87 \citep{kpc08}. Nevertheless, we decided not to include this information as this is not an output from the current observations and because we wanted to test whether our simulations will converge to values close to 0.87. 

For the radius of the loop as observed in Figure \ref{raftery_EM_T_t} from \citet{raftery08}, we assume that the ratio $R=\frac{r'}{L'}$ is probably $1/6$. It is almost certainly no more than $1/2$; so we give the ratio a $99.73\%$ probability to be inside the interval $[0,0.5]$. For this reason we have $R \sim \mathcal{G}(6.03,0.03)$. Since $r'=R \times L'$ and $R, L'$ have known prior distributions, we can simulate the prior distribution of $r'$ to be $\mathcal{G}(5.23,0.12)$, which will give $95\%$ probability between $[0.20, 1.24]$ with mode at $0.49$. 

Since there is no prior information regarding the $\alpha$ parameter, it would be preferable to assign an uninformative prior just as with $B'$. However, due to the fact that it is the ``important'' parameter of the analysis (if we want to test the hypotheses $H_i, i=0, \ldots, 3$), an improper prior would lead to Bartlett's paradox \citep{lind57,bar57}\footnote{In brief, Bartlett's paradox states that the less informative the prior of the ``important'' parameters is, the more the simpler models will be favoured. On the other hand, if the prior of the ``important'' parameters is very informative, then the more complex model will be favoured. This implies that the model selection method can be very sensitive to the prior information, which is somewhat bizarre.}. Therefore, for sensitivity reasons, results from an informative prior with $50\%$ probability for $\alpha>1$ and $50\%$ probability for $\alpha<1$ will be presented as well. For the $H_0$ hypothesis ($\alpha \neq 1$), this probability density function will be of the form: 
%However, the fact that it is the ``important'' parameter of the analysis (if we want to test the hypotheses $H_i, i=0, \ldots, 3$) it has been given an informative prior, otherwise this would lead to Bartlett's paradox \citep{lind57,bar57}\footnote{In brief, Bartlett's paradox states that the less informative the prior of the ``important'' parameters is, the more the simpler models will be favoured. On the other hand, if the prior of the ``important'' parameters is very informative, then the more complex model will be favoured. This implies that the model selection method can be very sensitive to the prior information, which is somewhat bizarre. 
%In Section \ref{sec:MCM} two model selection methods are presented: one that is very sensitive to the prior information (Bayes factor) and another that is not as sensitive (posterior deviances).
%}. Thus, a probability density function that will give a $50\%$ probability for $\alpha<1$ and a $50\%$ probability for $\alpha>1$ was chosen. For the $H_0$ hypothesis ($\alpha \neq 1$), this probability density function will be of the form: 
\begin{displaymath}
\pi(\alpha) = \left\{ 
\begin{array}{ll}
0.5, \quad 0<\alpha<1 \\
0.5\exp(1-\alpha), \quad \alpha>1;  
\end{array}
\right.
\end{displaymath}
for the $H_2$ hypothesis ($\alpha>1$), $\pi(\alpha)=\exp(1-\alpha)$; and for the $H_3$ hypothesis ($\alpha<1$), $\alpha \sim \mathcal{U}(0,1)$. As far as the true time is concerned we can assume that it is normally distributed about the observed time with standard deviation obtained from the time error bars. 

Our last assumption involves the initial values of the temperature, density and emission measure. For this, we assume the pre-flare conditions to have an upper limit of 0.5~MK, $6~\times 10^7$~cm$^{-3}$ and $4~\times~10^{43}$~cm$^{-3}$ respectively \citep{raftery08}. Since the non-thermal heating rate is almost zero in the beginning, then because of Equation (\ref{alpha_param_hest_rate}), the dominant heat input will be the background heating rate. Hence, these upper limits can indirectly serve as an upper limit to the background heating rate.

\section{Model Comparison Methods}
\label{sec:MCM}

\subsection{Direct prior inclusion}

Model comparison methods using Bayesian statistics are reviewed in great detail in \citet{clyde07}. An excellent description of how Bayes factor is applied in several problems can be found in \citet{kr95} and \citet{rafb96}. The key value for estimating the Bayes factor is to calculate the marginal density function: 
\begin{equation*}
p(\mathbf{D}|M_k) = \int_{\mathbf{P}} {p(\mathbf{D}|\mathbf{P}) \pi(\mathbf{P}) d\mathbf{P}}, 
\label{marginal}
\end{equation*}
where $M_k$ is the hypothesis (model) under consideration, $p(\mathbf{D}|\mathbf{P})$ is the likelihood function and $\pi(\mathbf{P})$ is the prior distribution. Roughly speaking, the model with the largest marginal density will provide evidence of its favour among a given class of models. Essentially, the marginal distribution is nothing more than the belief that we have about the data after we have integrated (averaged) over all the parameters, for the specific model under consideration. Then, the Bayes factor for testing two hypotheses ($M_k$ and $M_l$) is given by: 
\begin{equation}
\label{Bayes_fact}
BF_{k,l} = \frac{p(\mathbf{D}|M_k)}{p(\mathbf{D}|M_l)}. 
\end{equation}

From Equation~(\ref{Bayes_fact}) it is clear that if $BF_{k,l} \gg 1$ there is more evidence in favour the $M_k$ model, if $BF_{k,l} \ll 1$ there is more evidence in favour the $M_l$ model, whereas if $BF_{k,l} \approx 1$ the data do not favour either model \citep[but see][for more information]{kr95}. Thus, the challenge here is to calculate the marginal densities. Since they are very difficult to calculate analytically, we will turn to numerical methods. In this paper a two-stage MCMC sampler \citep{mira01} is employed. Laplace method and importance sampling methods for marginal likelihood estimation as well as the MCMC sampler with transformation of the parameter set for better convergence are described in detail in \citet{adamakis09}.

\subsection{Indirect prior inclusion}

Although Bayes factor has an intuitive interpretation, it has two major drawbacks: (\emph{i}) it is heavily dependent on the choice of the prior distribution for the important parameters, and, (\emph{ii}) accurate and efficient computation of the marginal likelihood can be difficult. Attempts have been made to mitigate the first issue by using the intrinsic Bayes factor \citep{bp96, bp98} or the fractional Bayes factor \citep{ohagan95}. According to these methods, the reasearcher should have large enought data, so that the sample can be split into a training set to estimate the posterior distribution of the parameters, which in turn will be used as the prior distribution in the remaining sample. However, the Bayes factor still depends to the choice of the training set. For this reason, one has to take into account each possible training set and calculate the new Bayes factor as the average of all the Bayes factors. The application of this was not feasible for the present study, where we have only 13 data-points: 7 temperature measures and 6 emission measures. Therefore, in this paper, results from information criteria and posterior deviances will be presented for comparison. 

\subsubsection{Information criteria}

The Akaike Information Criterion (AIC), proposed by \citet{ak74}, and the Bayesian Information Criterion (BIC), proposed by \citet{sc78} are also employed in the current study. AIC propose to choose the model that minimises:
\begin{equation*}
\label{AIC_equation}
\mbox{AIC} = -2(\textrm{log maximised likelihood}) + 2 \lambda,
\end{equation*}
whereas the latter chooses the model that minimises:
\begin{equation*}
\label{BIC_equation}
\mbox{BIC} = -2(\textrm{log maximised likelihood}) + \lambda \log n ,
\end{equation*}
where $\lambda$ is the number of the parameters and $n$ is the number of the data-points. The difference between two BICs: 
\begin{equation*}
B_{k,l} = \exp \left[ - \frac{1}{2} \left( BIC_k - BIC_l  \right)   \right]
\end{equation*}
can be viewed as an approximation to the Bayes factor without the researcher's comformable choice of priors. However, one has to come to terms with the fact that, indirectly, a prior similar to the Jeffreys prior is incorporated \citep{kw95}. The most important difference between AIC and BIC is that AIC was designed to find the model that produces estimates of the density which is close on average to the true density\footnote{Close is measured by the Kullback-Leibler distance.}, whereas BIC was designed to find the most probable model given the data \citep{wasserman00}.

\subsubsection{Posterior deviance}

Another promising method for comparing different models can be to only use the posterior deviance distribution of each competing model \citep{aitkin09, liu08}. According to this method, suppose that $p_k(\mathbf{P}^{[t]}|\mathbf{D})$ is the $t$th independent MCMC draw of the posterior likelihood function given the data $\mathbf{D}$ and model $k$. Furthermore, assume that the $t$th independent MCMC draw of the posterior deviance function for model $k$ is $D_k(\mathbf{P}^{[t]}|\mathbf{D}) = -2 \log{p_k(\mathbf{P}^{[t]}|\mathbf{D})}$, where $\log{(\cdot)}$ is the natural logarithm function. From this, the distribution of the difference between two model deviances can be calculated as $D_{k,l} = D_k - D_l$. Therefore, the posterior deviance difference distribution can be used to calculate the probability $P[D_{k,l}>\beta|\mathbf{D}]$. For large numbers of $\beta$, the higher this probability the stronger evidence there is in favour of model $l$ against model $k$. As is always the case with model selection, subjectivity comes before objectivity. That is, in order to objectively choose a model, we need to subjectively specify a rule that will lead us to this choice. We present here three different ways that can help a researcher to use posterior deviances for model selection. 

\begin{enumerate}
\item  The choice of the value of $\beta$ can vary for different applications. For example, for $\beta=0$ our value of interest is translated to the probability of model $k$ having just a higher deviance than model $l$. In order to make an association with Classical statistics, the difference between the deviances of two nested models is approximately a chi-square distribution with $df$ degrees of freedom, where $df=\nu_l - \nu_k$, the difference between the number of parameters estimated. Then the difference between the two deviances is compared with $\chi_{1-\alpha;df}$, the value that leaves probability $\alpha$ to the right tail of a chi-square distribution with $df$ degrees of freedom. For example, if $\alpha=0.05$ and $df=1$ then $\chi_{0.95;1}=3.84$. In this case we can choose $\beta=\chi_{1-\alpha;df}$. \citet{aitkin09} use $\beta=4.4$ as this gives a posterior probability of $0.9$ for model $l$, under the assumption of equal prior probabilities on each model. Their suggestion is that if $P[D_{k,l}>4.4|\mathbf{D}] > 0.9$ then there is quite strong evidence in favour of model $l$ against model $k$. 
\item  One can also try to find the value of $\beta$ that gives: 
\begin{equation}
\label{eq:beta}
P[D_{k,l}>\beta|\mathbf{D}]=0.50, \quad \beta~\ge~0.
\end{equation} 
If we adopt the table in \citet{kr95}, then we will end up with Table \ref{post_table} for the range of $\beta$. It is worth noting that these numbers are driven more from intuition, rather than a scientific justification. 
\item   Another way to quantify our beliefs is to calculate the value of $\gamma$ that gives 
\begin{equation}
\label{eq:gamma}
P[D_{k,l}>0|\mathbf{D}]=\gamma, \quad \gamma~\ge~0.50. 
\end{equation}
\end{enumerate}

\begin{figure}
   \centering
   \includegraphics[scale=0.65]{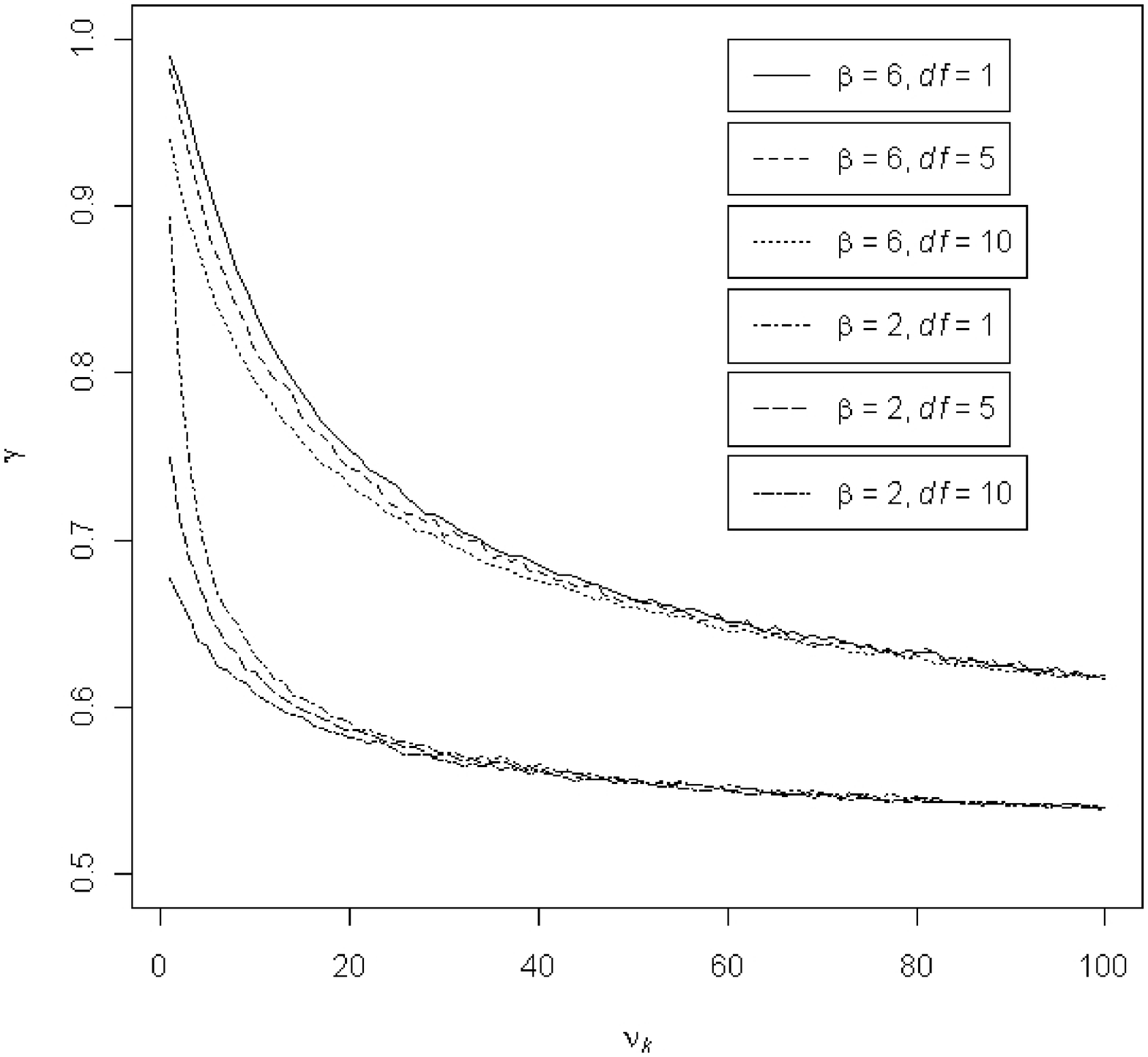} 
   \caption{$\gamma$ as a function of $\beta, df$ and $\nu_k$. If we assume that the values of $\beta$ that satisfy Equation~(\ref{eq:beta}) can be binned as in Table \ref{post_table} in order to quantify our beliefs, then for a given value of $\beta$ the value of $\gamma$ that satisfies Equation~(\ref{eq:gamma}) varies according to $df$ and $\nu_k$. }
   \label{fig:gamma_pic}
\end{figure}

\paragraph{Comparison of the three methods}
There appears to be a connection between $\beta, \gamma, df$ and $\nu_k$. That is, for fixed values of $\beta$ and $df$ the more parameters in a model, the smaller the value of $\gamma$, in order to obtain the same information. For example, if we adopt the approximation $D_k(\mathbf{P}|\mathbf{D}) \approx D(\widehat{\mathbf{P}}) + \chi_{\nu_k}$, where $D(\widehat{\mathbf{P}})$ is the frequentist deviance and $\chi_{\nu_k}$ has a chi-square distribution with $\nu_k$ degrees of freedom \citep{spiegelhalter02}, and also $\beta=2, df=2$ then for $\nu_k=1$ we get $\gamma=0.86$, whereas for $\nu_k=10$, we get $\gamma=0.63$. Figure \ref{fig:gamma_pic} depicts how $\gamma$ varies with $\nu_k$ for $df=\{1,5,10\}$ and $\beta=\{2,6\}$. On the other hand, if we bin the values of $\gamma$, then $\beta$ will vary according to $\gamma, df$ and $\nu_k$. This result implies that in order to quantify our belief one should bin the values of $\beta$ or $\gamma$ but not both. 

In Section \ref{results_raftery} results for Bayes factors, AIC, BIC and posterior deviances are presented. Regarding the posterior deviances in particular, although we show results from the above three methods, we favour the second method, \emph{i.e.} comparing the value of $\beta$ that satisfies Equation (\ref{eq:beta}) with the values in Table \ref{post_table}.

\begin{table}
\begin{center}
\caption{Rule of thumb for quantifying our belief when comparing between two models. The values of $\beta$ satisfy Equation~(\ref{eq:beta}).}
\label{post_table}
\begin{tabular}{cc}     % define the column alignment
                           % l: left, c: center, r: right
  \hline\hline                   % horizontal line
  $\beta$ & Evidence \\
  \hline\hline
 0--2 & Not worth more than a bare mention \\
 2--6 & Positive \\
 6--10 & Strong \\
 $>10$ & Very strong \\
  \hline\hline
\end{tabular}
\end{center}
\end{table}

\section{Results}
\label{results_raftery}

The aim of this analysis is to address the following questions: 
\begin{enumerate}
\item Which of the Half Gaussian and Full Gaussian functions fits better to the data we have? 
\item Which of the thermal and non-thermal heat fluxes is dominant? 
\item Should the $c_i$ values remain fixed or should they be free parameters? 
\end{enumerate}

\begin{figure}[hbtp]
  \begin{center}
    \epsfig{figure=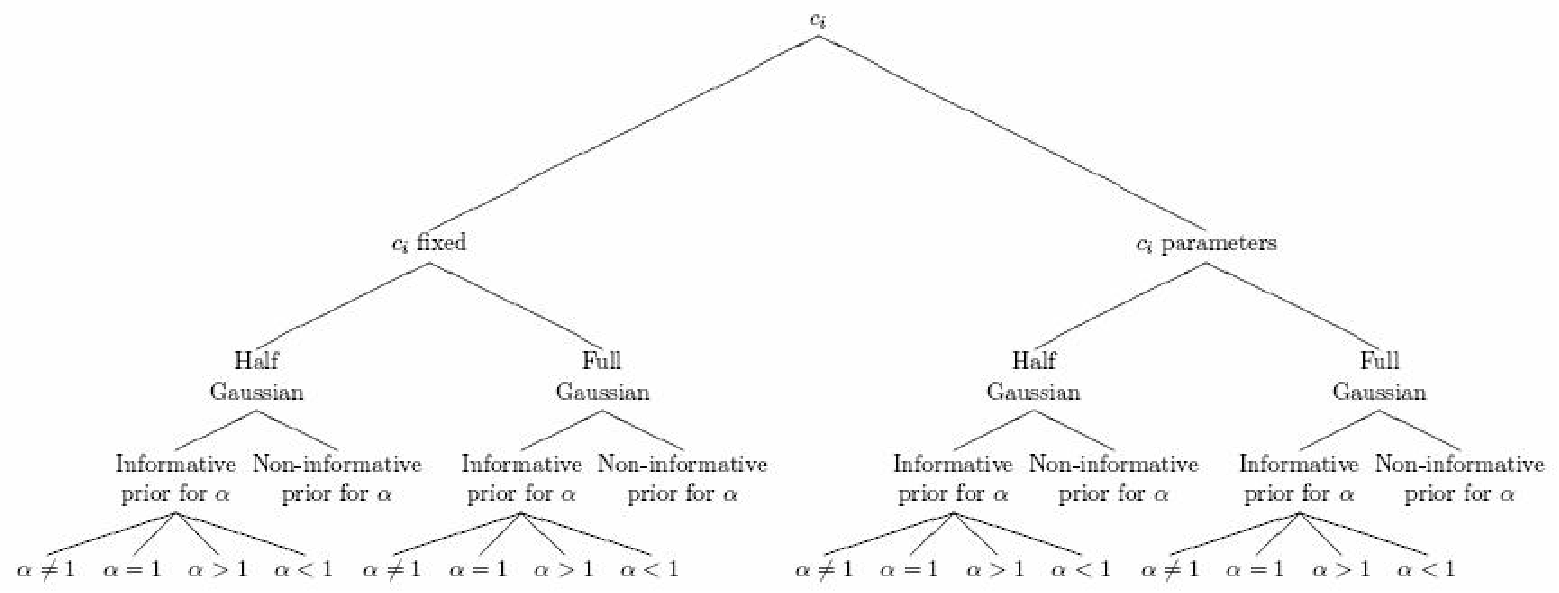, angle=90, scale=0.90}
    \caption{Tree diagram that depicts all the different hypotheses we have used. The purpose of using these hypotheses is to compare them and select the one that best describes the data-set we have.}
    \label{fig_rotate}
  \end{center}
\end{figure}

The second question is the most important regarding the physics of the system. The other two questions are associated more with the statistical analysis. Nevertheless, they can indirectly affect which heat flux is most dominant because they contain relevant parameters. The estimations of the marginal densities of the different models that are shown in Figure \ref{fig_rotate} are presented in Table \ref{c_i fixed:BF}. For all the statistical models we have assumed that the mean energy of the accelerated (non-thermal) electrons impinging the chromosphere is $\mathcal{E}=15$~keV. Different values of $\mathcal{E}$ have a minor effect on the temporal profiles.

\begin{table}
\caption{Model selection criteria according to marginal densities, AIC and BIC produced under several hypotheses using the EBTEL model to compare with temperature and emission measure profiles \citep{raftery08}. This table produces results for both $c_i$ fixed and free parameters for comparison tests. The logarithmic marginal densities are estimated using: 1: Laplace method with posterior covariance matrix, 2: Laplace method with robust posterior covariance matrix, 3: Importance sampling estimation with the probability density from the first stage of the two-stage sampler as the additional probability density. See \citet{adamakis08} for more information about these estimations.}
\label{c_i fixed:BF}
\begin{center}
\begin{tabular}{c|c|c|c|c|c|c|c|c}     % define the column alignment
                           % l: left, c: center, r: right
  \hline\hline                   % horizontal line%

  \multicolumn{3}{c}{} & & \multicolumn{3}{c|}{Log-marginal densities} & AIC & BIC  \\
%   \multicolumn{3}{c}{} & AIC & \multicolumn{3}{c}{Log-marginal densities} & AIC & BIC  \\

  \hline

%  \multicolumn{3}{c}{} & & 1 & 2 & 3 & & \\
  $c_i$ & Heat & $\alpha$ prior & Models & 1 & 2 & 3 & & \\
  \hline\hline
& &  & $M_1: \alpha \ne 1$ & -796.43 & -797.78 & -797.66 & 1589.83 & 1597.74 \\
& &  & $M_2: \alpha=1$ & -797.38 & -798.19 & -797.37 & 1587.22 & 1594.57 \\
&\raisebox{0.0ex}{HG} & \raisebox{1.0ex}{Informative} & $M_3: \alpha > 1$ & -797.20 & -798.29 & -797.53 & 1589.83 & 1597.74 \\
& &  & $M_4: \alpha < 1$ & -797.34 & -798.54 & -797.99 & 1590.17 & 1598.08 \\
\cline{3-9}
& & Non-informative & $M_5: \alpha \ne 1$ & -792.88 & -793.68 & -795.95 & 1589.83 & 1597.74 \\ 
\cline{2-9}
\raisebox{1.0ex}{fixed} & &  & $M_6: \alpha \ne 1$ & -793.82 & -795.28 & -794.84 & 1582.32 & 1590.23 \\
& &  & $M_7: \alpha=1$ & -794.45 & -795.39 & -794.40 & 1581.13 & 1588.48 \\
&\raisebox{0.0ex}{FG}       & \raisebox{1.0ex}{Informative} & $M_8: \alpha > 1$ & -793.87 & -794.93 & -794.42 & 1582.32 & 1590.23 \\
& &  & $M_9: \alpha < 1$ & -794.17 & -795.31 & -794.73 & 1582.73 & 1590.64 \\
\cline{3-9}
& & Non-informative & $M_{10}: \alpha \ne 1$ & -789.07 & -790.48 & -792.21 & 1582.32 & 1590.23 \\
  \hline
  \hline
& &  & $M_{11}: \alpha \ne 1$ & -793.26 & -795.98 & -798.47 & 1573.52 & 1583.12 \\
& &  & $M_{12}: \alpha=1$ & -793.36 & -796.26 & -798.06 & 1571.88 & 1580.92 \\
&\raisebox{0.0ex}{HG} & \raisebox{1.0ex}{Informative} & $M_{13}: \alpha > 1$ & -794.07 & -795.91 & -797.96 & 1573.52 & 1583.12 \\
& &  & $M_{14}: \alpha < 1$ & -793.23 & -796.60 & -797.62 & 1573.98 & 1583.58 \\
\cline{3-9}
& & Non-informative & $M_{15}: \alpha \ne 1$ & -789.65 & -792.36 & -796.98 & 1573.52 & 1583.12 \\ 
\cline{2-9}
%\parbox{2cm}{free \\ par.} 
\raisebox{0.0ex}{free par.} 
& &  & $M_{16}: \alpha \ne 1$ & -791.06 & -793.25 & -794.28 & 1568.66 & 1578.26 \\
& & {Informative} & \raisebox{1.0ex}{$M_{17}: \alpha=1$} & \raisebox{1.0ex}{-790.26} & \raisebox{1.0ex}{-792.38} & \raisebox{1.0ex}{-794.39} & \raisebox{1.0ex}{1567.62} & \raisebox{1.0ex}{1576.66} \\
&\raisebox{0.0ex}{FG}  &  & \raisebox{1.0ex}{$M_{18}: \alpha > 1$} & \raisebox{1.0ex}{-789.99} & \raisebox{1.0ex}{-792.37} & \raisebox{1.0ex}{-794.32} & \raisebox{1.0ex}{1568.66} & \raisebox{1.0ex}{1578.26} \\
& &  & \raisebox{1.0ex}{$M_{19}: \alpha < 1$} & \raisebox{1.0ex}{-791.48} & \raisebox{1.0ex}{-793.32} & \raisebox{1.0ex}{-794.71} & \raisebox{1.0ex}{1569.05} & \raisebox{1.0ex}{1578.65} \\
\cline{3-9}
& & Non-informative & $M_{20}: \alpha \ne 1$ & -786.85 & -788.49 & -792.02 & 1568.66 & 1578.26 \\
  \hline\hline
\end{tabular}
\end{center}
\end{table}

\begin{table}
\caption{Posterior distribution information for the $\alpha$ parameter. Results are split according to $c_i$ (fixed/parameters), heating flux (Half Gaussian/Full Gaussian) and $\alpha$ prior (informative/non-informative).}
\label{post_interv}
\begin{center}
\begin{tabular}{|c|c|c|c|c|c|c|}     % define the column alignment
                           % l: left, c: center, r: right
  \hline\hline                   % horizontal line%

  \multicolumn{3}{|c|}{} & \multicolumn{3}{|c|}{Quantiles} & \multirow{2}{*}{$P(\alpha>1)$} \\ 

  \cline{1-6}
%  \multicolumn{3}{c}{} & & 1 & 2 & 3 & & \\
  $c_i$ & Heat & $\alpha$ prior & $2.5\%$ & $50\%$ & $97.5\%$ & \\
  \hline
\multirow{4}{*}{fixed} & \multirow{2}{*}{HG} & Informative ($M_1$) & 0.16 & 0.91 & 3.11 & 0.44 \\
\cline{3-7}
& & Non-informative ($M_5$) & 0.34 & 8.10 & 48.47 & 0.90 \\ 
\cline{2-7}
 & \multirow{2}{*}{FG} & Informative ($M_6$) & 0.22 & 1.10 & 3.70 & 0.56 \\
\cline{3-7}
& & Non-informative ($M_{10}$) & 1.30 & 5.74 & 101.30 & $\sim~1$ \\ 
\cline{2-7}
  \hline\hline
\multirow{4}{*}{free par.} & \multirow{2}{*}{HG} & Informative ($M_{11}$) & 0.07 & 0.90 & 2.93 & 0.45 \\
\cline{3-7}
& & Non-informative ($M_{15}$) & 0.42 & 11.63 & 32.31 & 0.91 \\ 
\cline{2-7}
 & \multirow{2}{*}{FG} & Informative ($M_{16}$) & 0.13 & 1.13 & 4.08 & 0.57 \\
\cline{3-7}
& & Non-informative ($M_{20}$) & 1.13 & 23.48 & 47.09 & 0.98 \\ 
\cline{2-7}
  \hline\hline
\end{tabular}
\end{center}
\end{table}

Our decision about $\alpha$ will be based on AIC, BIC and the posterior deviance. Bayes factors are presented as well for both proper and improper priors for $\alpha$. However, since there is no real prior information for $\alpha$, Bayes factors should be used for guidance and not for decisioning. 
%Since we do not have any prior information for the $\alpha$ parameter and we have decided to give an informative prior distribution in order to avoid Bartlett's paradox, we have undertaken the same analysis with a non-informative prior for $\alpha$. The reason for this is because we wanted to check the sensitivity of the analysis regarding the prior restriction of the $\alpha$ parameter. Thus, we use the results obtained from Table \ref{c_i fixed:BF} with non-informative prior for $\alpha$ to approach Questions 1 and 3 and the results obtained with informative priors for $\alpha$ to decide which heating mechanism (thermal or non-thermal) prevails. 

\subsection{$c_i$ Fixed}
\label{c_i fixed}

\subsubsection{Model selection}
\label{mod_sel_fix}

\begin{figure}
   \centering
   \includegraphics[scale=0.45]{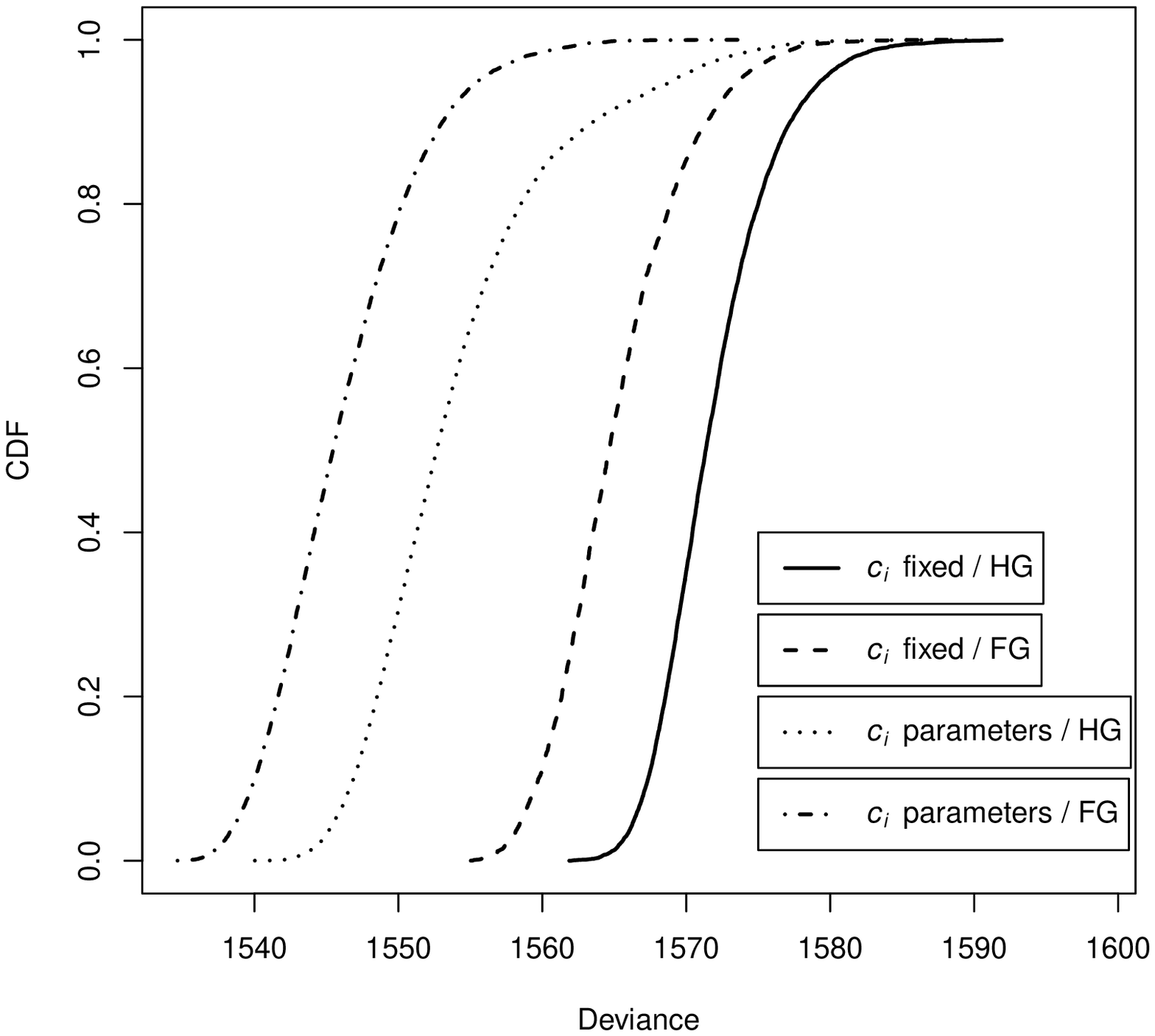} \hfil
   \includegraphics[scale=0.45]{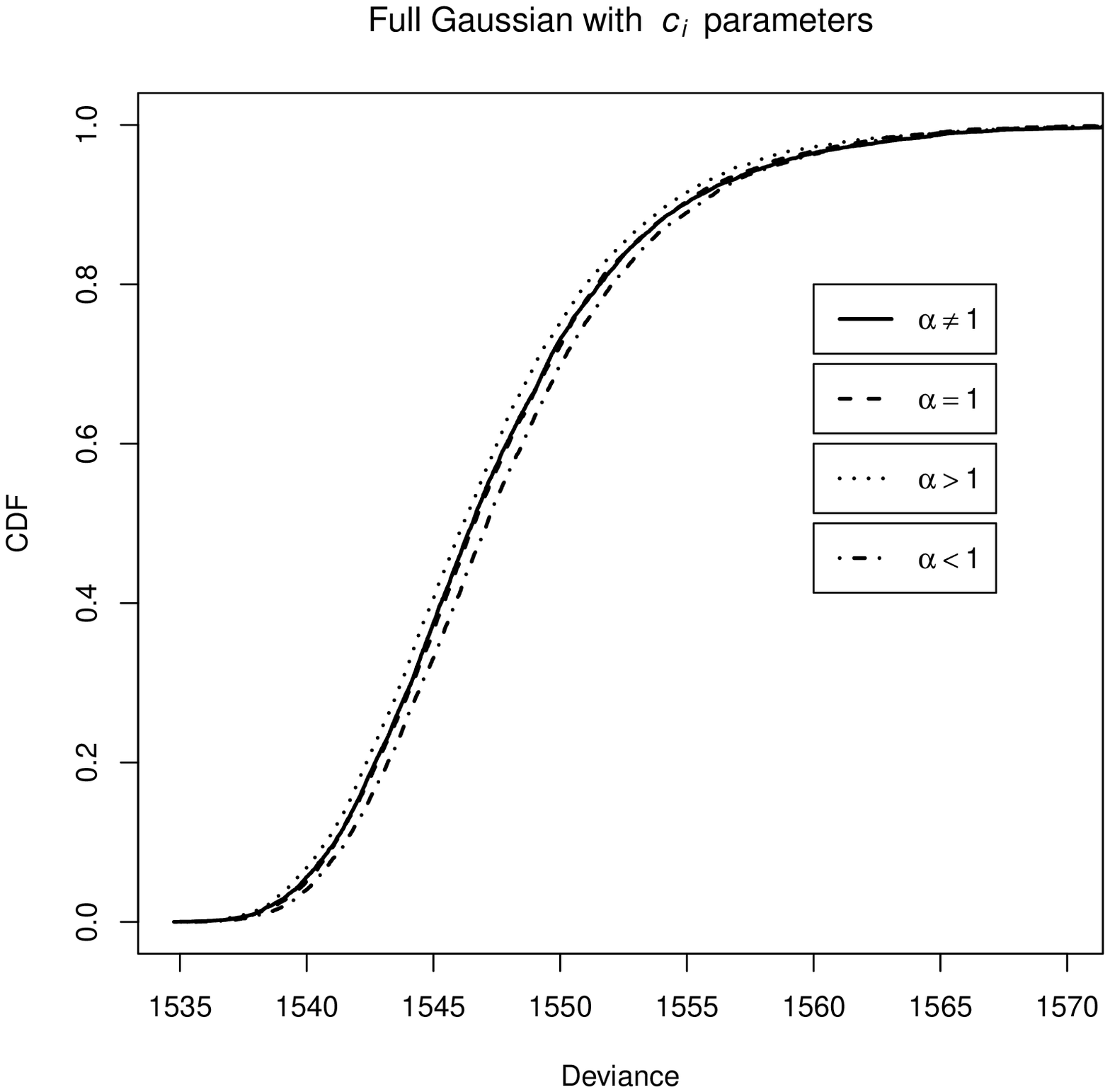} \hfil 
   \caption{\emph{Left}: Cumulative distribution functions for posterior deviance for models with $c_i$ fixed and Half Gaussian (solid), $c_i$ fixed and Full Gaussian (dashed), $c_i$ parameters and Half Gaussian (dotted), $c_i$ parameters and Full Gaussian (dot-dashed). All curves are with non-informative prior for $\alpha$. The further to the left, the better the model. \emph{Right}: Cumulative distribution functions for posterior deviance for $c_i$ parameters and Full Gaussian are presented for a closer comparison. The dashed line ($\alpha=1$) over plots the solid line ($\alpha \neq 1$). The dotted line ($\alpha > 1$) provides a slight better fit than the dashed line and the dot-dashed line ($\alpha < 1$) provides a slight worst fit than the dashed line. However, all the hypotheses provide very close lines, an indication that distinction between these models will be very difficult.}
   \label{fig:max_like}
\end{figure}

\begin{figure}
   \centering
   \includegraphics[scale=0.38]{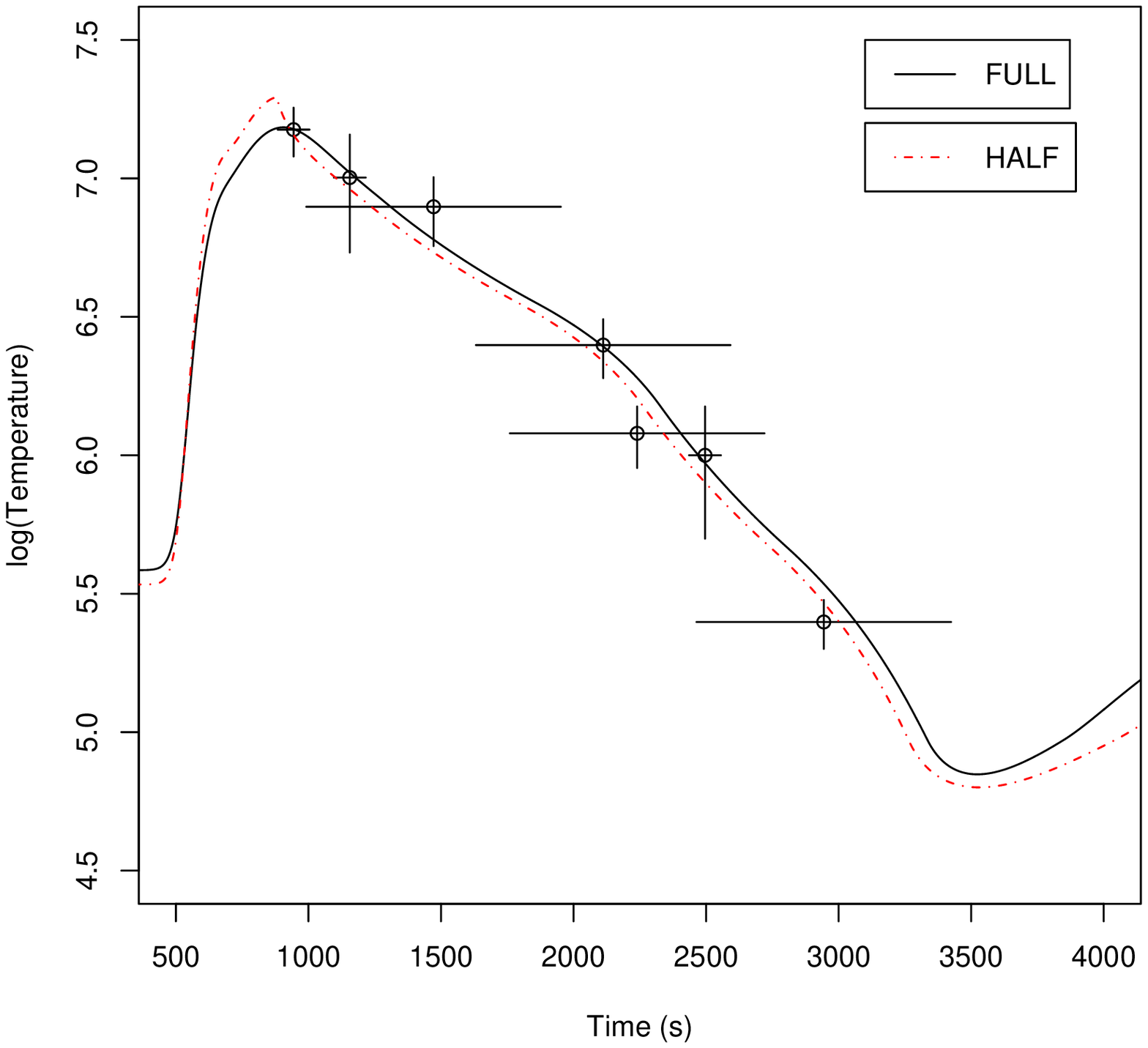}\hfill
   \includegraphics[scale=0.38]{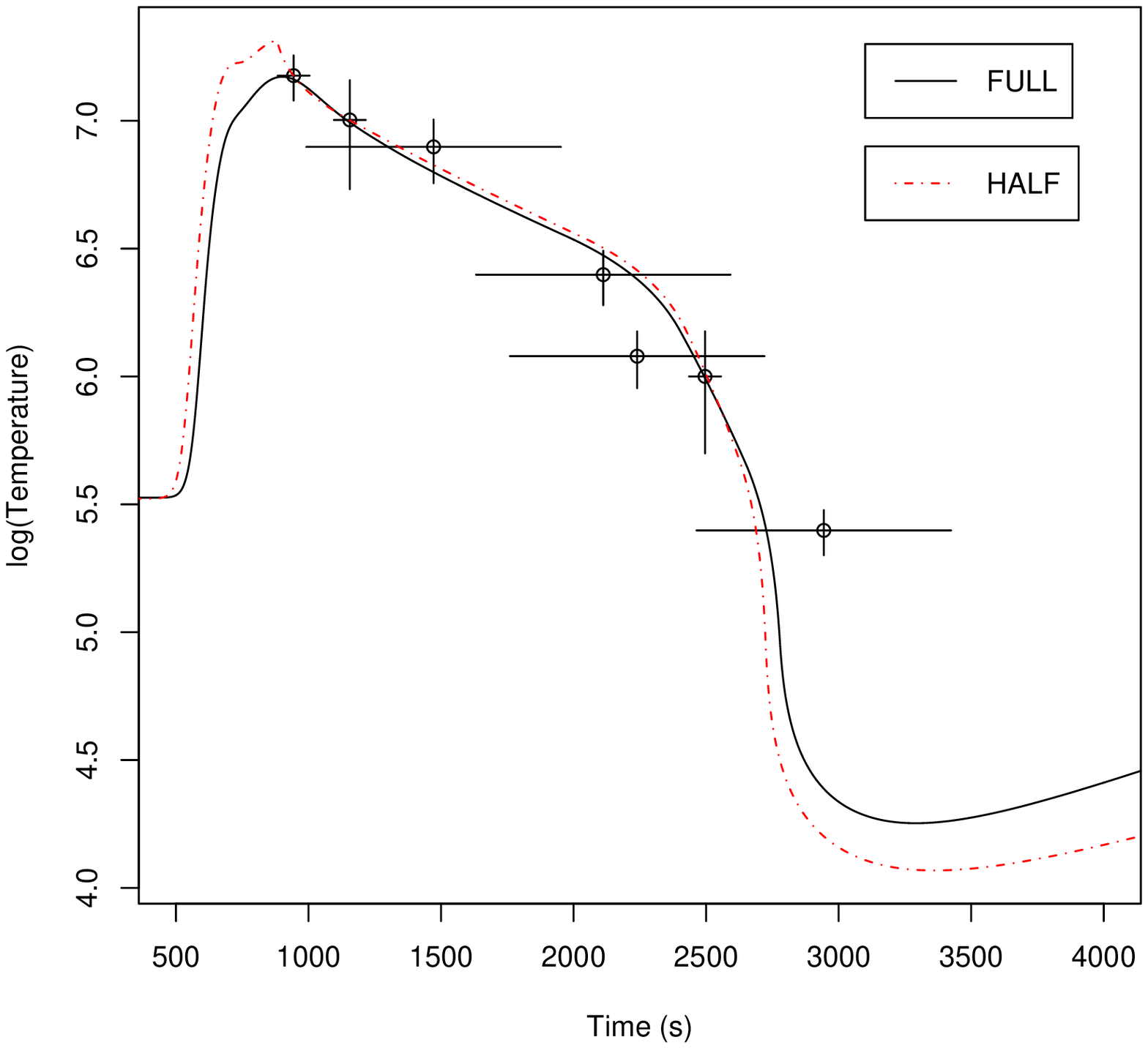}\hfill
   \includegraphics[scale=0.38]{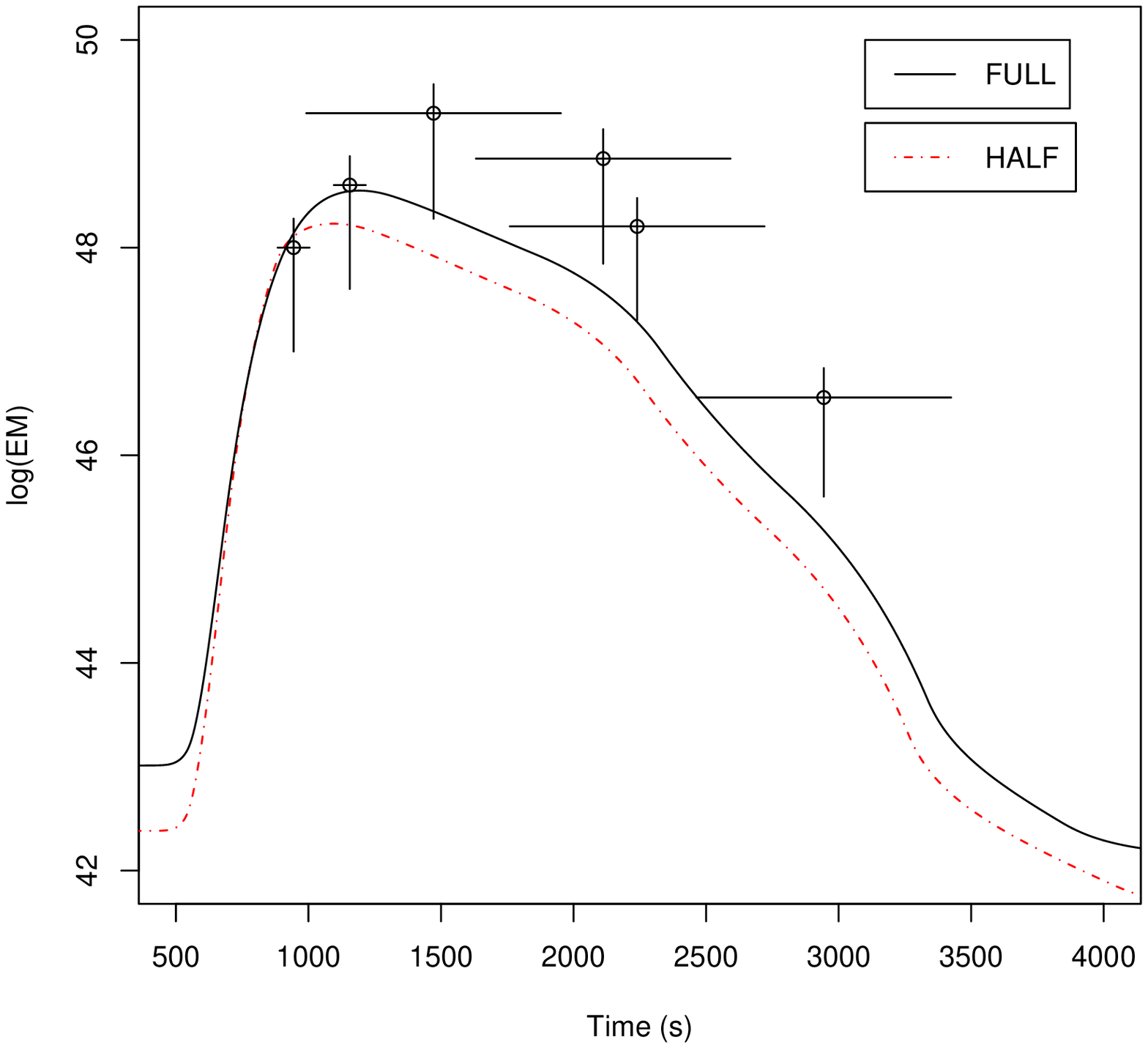}\hfill
   \includegraphics[scale=0.38]{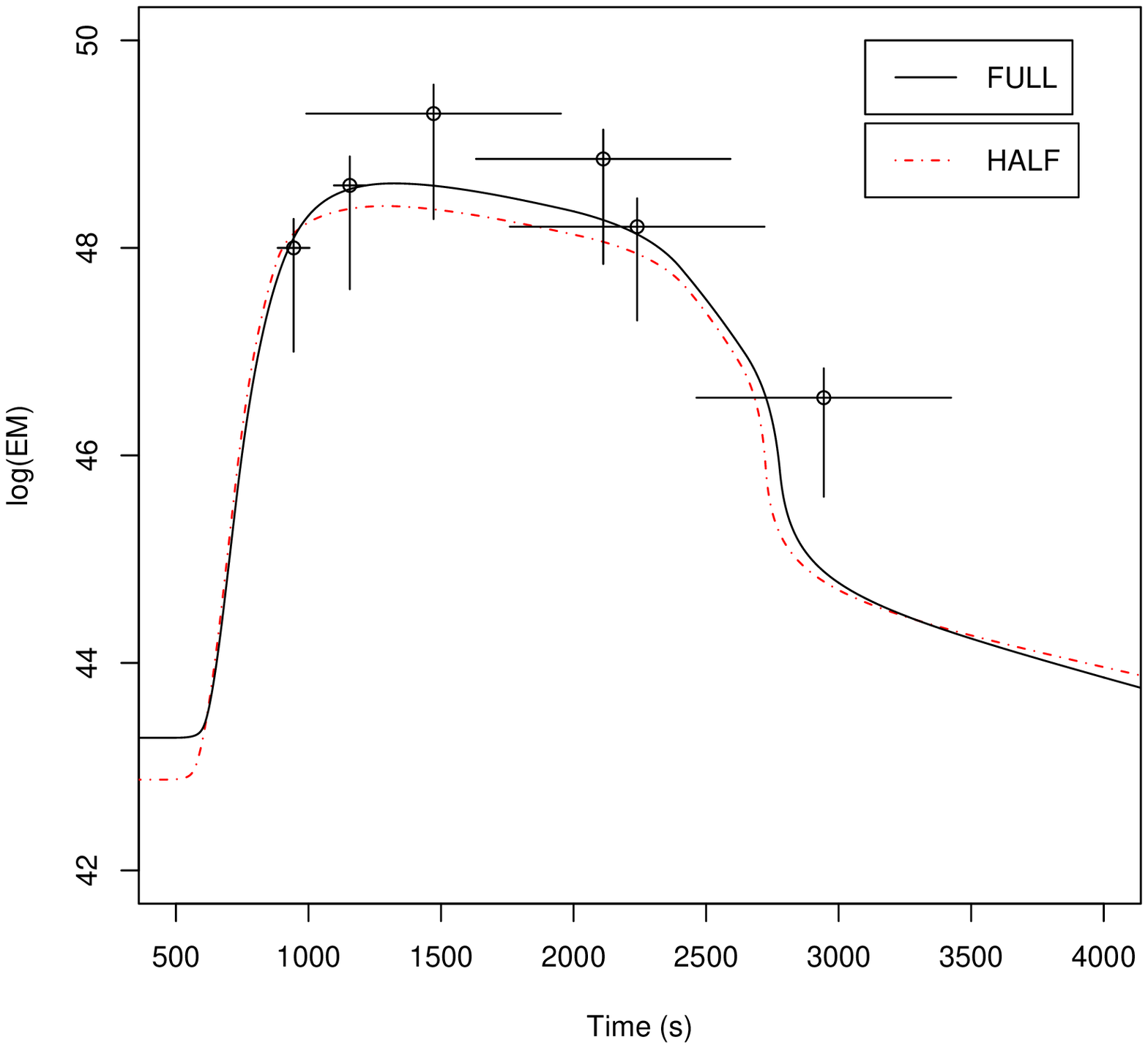}\hfill
   \includegraphics[scale=0.38]{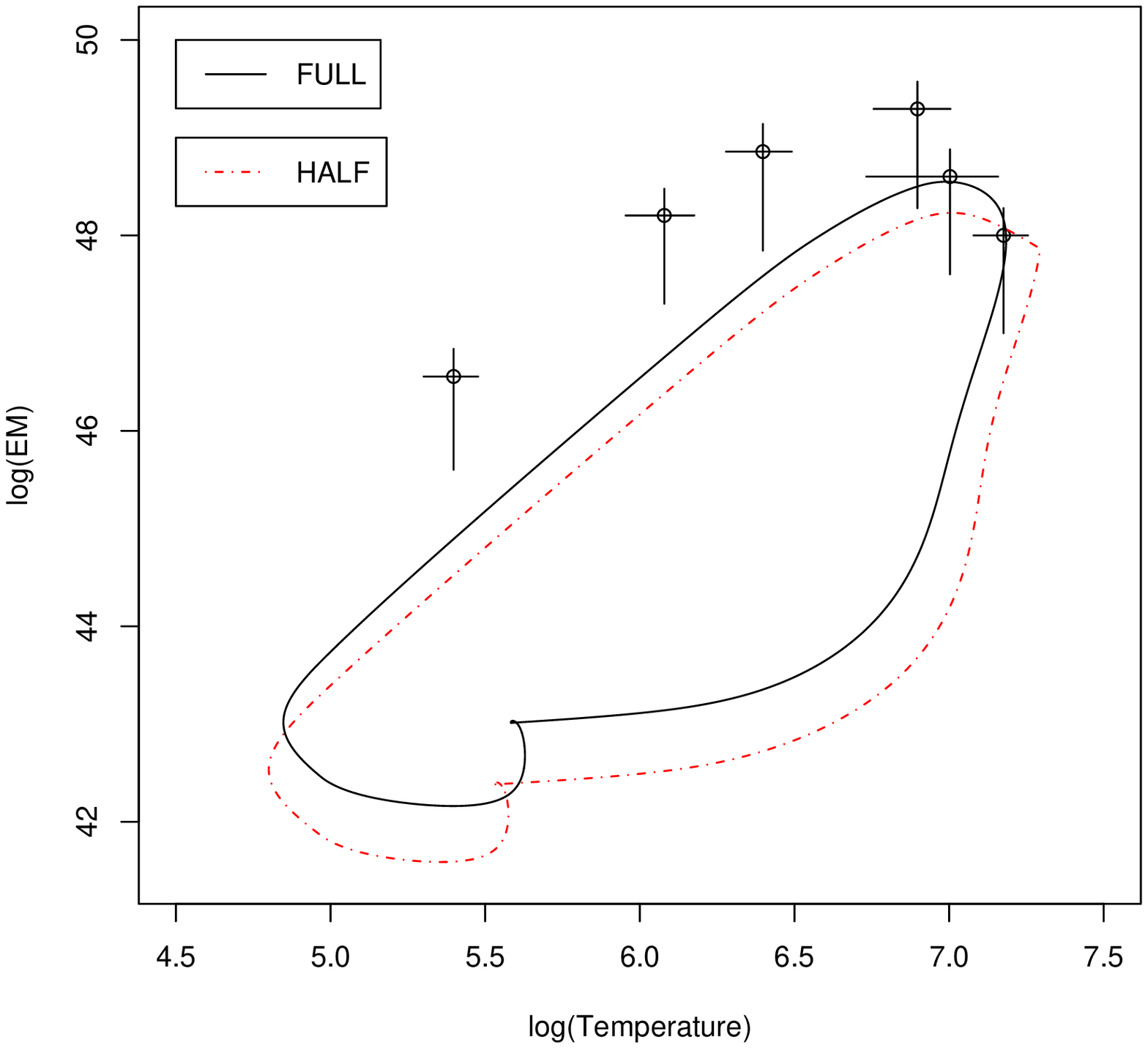}\hfill
   \includegraphics[scale=0.38]{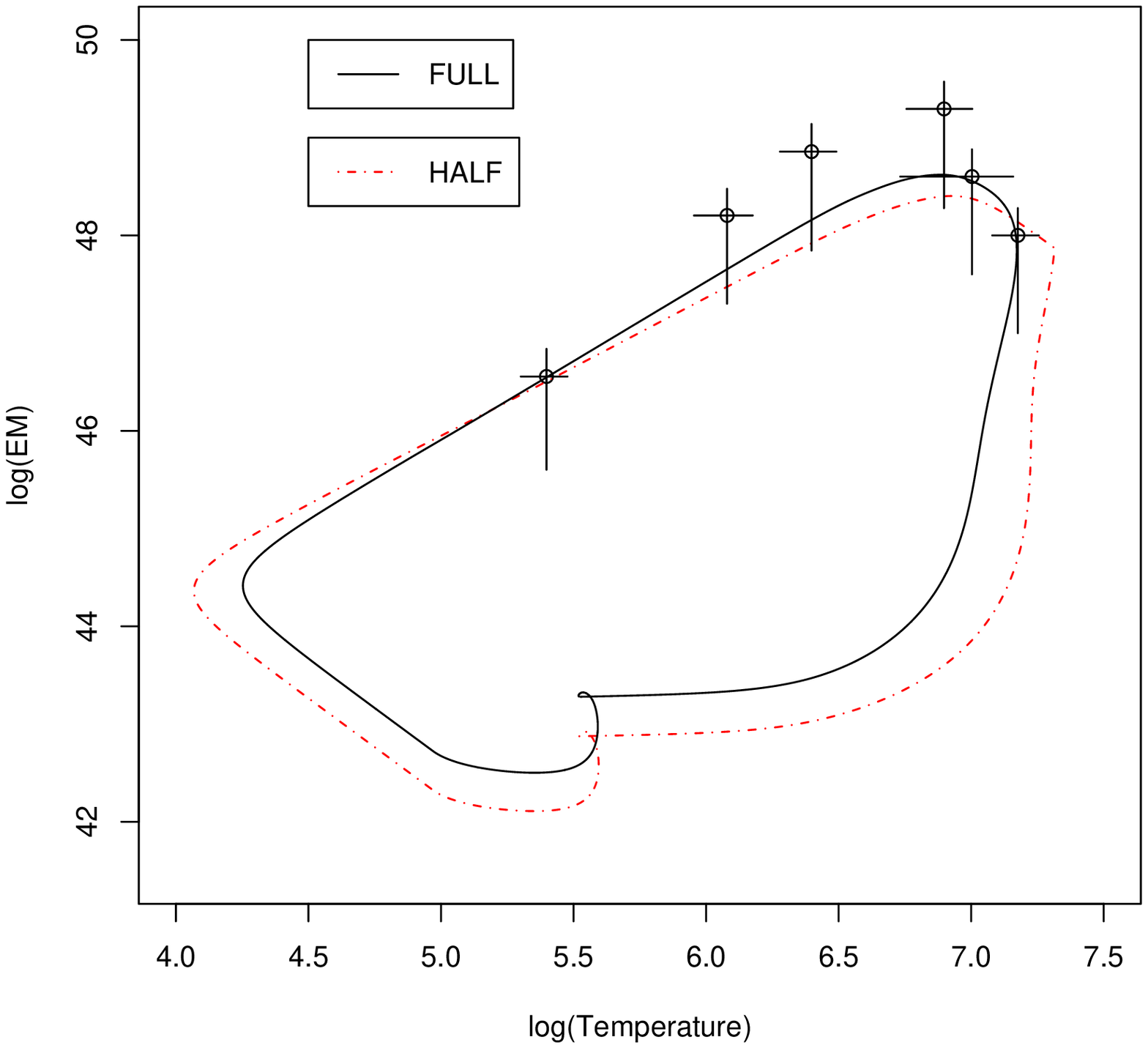}\hfill
   \caption{Temperature evolution ({\it first row}), emission measure evolution ({\it second row}) and temperature against emission measure ({\it third row}) using the EBTEL model that best reproduced the observations (maximised the posterior distributions). Solid lines represent the fit from the Full Gaussian function for the thermal and non-thermal heat inputs, whereas dashed lines represent the fit from the Half Gaussian function. For all the curves a {\bf non-informative} prior for $\alpha$ was employed. The {\it first column} depicts solutions using the $c_i$ fixed, whereas the {\it second column} depicts solutions from $c_i$ free parameters. Temperature is measured in K and emission measure is in $\mbox{cm}^{-3}$.}
   \label{fig1}
\end{figure}

The $c_i$ parameters are fixed at $c_1=4, c_2=0.87, c_3=0.72$ due to an acceptable overall agreement with 1D HD simulations \citep{raftery08,kpc08}. From Table \ref{c_i fixed:BF} and with a non-informative prior for $\alpha$ we can derive that the Bayes factor between the Full Gaussian and the Half Gaussian heating function is 42.10 using the Importance sampling estimator. This provides ``strong'' evidence in favour of the Full Gaussian, according to the table in \citet{kr95}. Similar decision is reached when comparing the marginal densities for informative $\alpha$ prior. 
Comparing the solid and the dashed lines at the left panel of Figure \ref{fig:max_like} we can conclude that Full Gaussian gives a better fit to the data. Also, $P[D_{5,10}>0|\mathbf{D}]=0.86$, $P[D_{5,10}>4.4|\mathbf{D}]=0.64$ and $P[D_{5,10}>6.3|\mathbf{D}]=0.50$. According to Table \ref{post_table} there is ``strong'' evidence in favour of the Full Gaussian. Furthermore, both information criteria prefer the Full Gaussian model. Finally, the first column of Figure \ref{fig1} shows the best fit of the parameters regarding the posterior distributions. Even by eye we can distinguish between the Full Gaussian and the Half Gaussian function. 

From all the above, we can conclude that there is enough evidence to support that the Full Gaussian function for the heating profiles is much more adequate than the Half Gaussian, at least for the particular data-set under analysis.

In order to distinguish between the thermal and non-thermal heat fluxes, we have calculated $P[D_{7,8}>0.6|\mathbf{D}]=0.50$, $P[D_{9,7}>0.6|\mathbf{D}]=0.50$ and $P[D_{9,8}>1.2|\mathbf{D}]=0.50$. Otherwise, $P[D_{7,8}>0|\mathbf{D}]=0.54$, $P[D_{9,7}>0|\mathbf{D}]=0.54$ and $P[D_{9,8}>0|\mathbf{D}]=0.58$. According to Table \ref{post_table} none of the models is ``worth more than a bare mention''. On the other hand, both information criteria choose the equal amounts of thermal and non-thermal heating model ($M_7$), although its AIC and BIC values are very close to the ones of $M_8$ and $M_9$. Furthermore, the marginal densities of $M_7, M_8, M_9$ are very similar to each other, something that supports the idea that the difference between the models is ``not worth more than a bare mention''. Finally, according to the quantiles of Table \ref{post_interv}, the $95\%$ credible interval with informative prior for $\alpha$ contains 1, whereas the $95\%$ credible interval with non-informative prior for $\alpha$ does not contain 1. In other words, $P(\alpha>1)$ is close to 0.50 with informative prior for $\alpha$, whereas it is close to 1 with non-informative prior for $\alpha$. 

%In order to distinguish between the thermal and non-thermal heat fluxes, let us focus on the results from Table \ref{c_i fixed:BF} with the informative prior for $\alpha$. Roughly speaking, the model with the largest marginal density will provide evidence of its favour among a given class of models. Using the Full Gaussian we conclude that the evidence we get is ``not worth more than a bare mention'' for any hypothesis ($\alpha>1$, $\alpha<1$, $\alpha=1$). We reach the same conclusion for the Half Gaussian as all the marginal probability densities are almost the same. 
%The posterior deviances from the top panel of Figure \ref{fig:max_like} are in agreement with this as all the lines are very close together. For example, we have calculated $P[D_{7,8}>0.6|\mathbf{D}]=0.50$, $P[D_{9,7}>0.6|\mathbf{D}]=0.50$ and $P[D_{9,8}>1.2|\mathbf{D}]=0.50$. Otherwise, $P[D_{7,8}>0|\mathbf{D}]=0.54$, $P[D_{9,7}>0|\mathbf{D}]=0.54$ and $P[D_{9,8}>0|\mathbf{D}]=0.58$. According to Table \ref{post_table} none of them is ``worth more than a bare mention''. 

In summary, it is very difficult to distinguish which of the two heating forms is more dominant. Better data are required to address this issue. Since the Full Gaussian is more preferable than the Half Gaussian we will choose to adopt the Full Gaussian results for estimating the parameters.

\subsubsection{Parameter estimation}
\label{par_est_fix}

The results of this analysis with non-informative prior for $\alpha$ can be viewed in Table \ref{c_i fix}. We estimate the mean of $L$ to be 29.20 Mm, which is close to what was used by \citet{raftery08} (\emph{i.e.} 30 Mm). The mean total non-thermal heat flux is calculated to be around $79.83 \times 10^9 $ ergs $\mbox{cm}^{-2}$, while the mean time for the peak of this function is 853.5 s after the beginning of the flare. For the $\alpha$ parameter a $95\%$ credible interval gives $[1.30, 101.30]$, whereas the probability $P(\alpha>1)=55.86\%$ with informative prior for $\alpha$. We expected the posterior probability of $\alpha>1$ to be close to 1/2, as the Bayes factor could not provide enough evidence in favour of the hypotheses $\alpha>1$ or $\alpha<1$ (Table \ref{c_i fixed:BF}). An interesting point to note here is that if we use non-informative prior for $\alpha$, we can conclude that since $\widehat{\alpha}=20.90$ then the $\alpha>1$ hypothesis is preferable. This is something that is misused in astrophysics and can lead to false conclusions. More discussion about this is presented in Section \ref{discussion}.

\begin{table}
\caption{Summary of the posterior inference for both $c_i$ fixed and free parameters with {\bf non-informative} prior for $\alpha$. Results steam from a Full Gaussian non-thermal heat flux profile.}
\label{c_i fix}
\begin{center}
\begin{tabular}{c|c|cccccc}     % define the column alignment
                           % l: left, c: center, r: right
  \hline\hline                % horizontal line
\multicolumn{1}{c}{} & & mean & mode & s.d. & $2.5\%$ & $50\%$ & $97.5\%$ \\
  \hline\hline
& $L'$       & 2.92 & 2.86 & 0.12 & 2.69 & 2.91 & 3.19 \\
& $\mathcal{F}'_{tot}$ & 79.83 & 44.72 & 62.96 & 3.77 & 75.73 & 179.88 \\
& $\mu$     & 853.5 & 881.1 & 31.6 & 763.0 & 863.8 & 883.1 \\
$c_i$ fixed & $\sigma_1$  & 101.8 & 94.2 & 26.9 & 54.4 & 99.6 & 154.6 \\
& $\alpha$  & 20.90 & 8.46 & 28.23 & 1.30 & 5.74 & 101.30 \\
& $B'$       & 0.19 & 0.25 & 0.11 & 0.01 & 0.20 & 0.36 \\
& $r'$       & 0.47 & 0.51 & 0.12 & 0.25 & 0.45 & 0.73 \\
  \hline
  \hline
& $L'$       & 3.16 & 2.80 & 0.43 & 2.43 & 3.13 & 4.12 \\
& $\mathcal{F}'_{tot}$ & 31.99 & 8.35 & 40.11 & 4.84 & 13.57 & 153.98 \\
& $\mu$     & 864.6 & 880.6 & 18.7 & 813.2 & 870.3 & 883.5 \\
& $\sigma_1$  & 92.6 & 82.0 & 24.2 & 51.8 & 90.5 & 146.2 \\
& $\alpha$  & 23.51 & 38.09 & 15.94 & 1.13 & 23.48 & 47.09 \\
\raisebox{0.5ex}{$c_i$ free parameters} & $B'$       & 0.12 & 0.21 & 0.07 & 0.01 & 0.17 & 0.24 \\
& $r'$       & 0.67 & 0.57 & 0.17 & 0.39 & 0.66 & 1.05 \\
& $c_1$       & 2.07 & 1.39 & 0.77 & 0.81 & 1.99 & 3.79 \\
& $c_2$       & 0.86 & 0.88 & 0.07 & 0.73 & 0.87 & 0.99 \\
& $c_3$       & 0.75 & 0.69 & 0.13 & 0.48 & 0.75 & 0.98 \\
  \hline\hline
\end{tabular}
\end{center}
\end{table}

%The first column of Figure \ref{fig1} shows the best fit of the parameters regarding the posterior distributions. Even by eye we can distinguish between the Full Gaussian and the Half Gaussian function. 
%As previously discussed, introducing a prior distribution for the $\alpha$ parameter, when there is no prior belief for this parameter, will create a bias. This is the price we have to pay in order to address Question 2. However, the prior we employed seems reasonable as it provides 1/2 probability for $0<\alpha<1$ and, subsequently, 1/2 probability for $\alpha>1$. In order to see how different the results are, both with and without prior information for the $\alpha$ parameter, then we compare Table \ref{c_i fix} with the respective table we get from the analysis where prior information for $\alpha$ is imported. From that we can conclude that all the parameters have almost the same posterior distributions, apart from the $\mathcal{F'}_{tot}$ parameter. Especially for this parameter, a $95 \%$ highest posterior density interval\footnote{A $95\%$ highest posterior density interval of a parameter leaves $95\%$ probability for the parameter to be inside this credible interval and any value of the posterior distribution that is inside this credible interval is higher than any value of the posterior distribution that is outside this credible interval \citep{carlin00,chen00,aitken04}.} will give $[51.51,544.56]$ with informative prior for $\alpha$ and $[2.11,171.90]$ with non-informative prior for $\alpha$. 

\subsection{$c_i$ Parameters}
\label{c_i parameters}

\subsubsection{Model selection}
\label{mod_sel_par}

We follow the same procedure as in Section \ref{mod_sel_fix} in order to derive the models that best describe the data. Regarding which of the two non-thermal heat fluxes is more dominant, there is no question that the Full Gaussian is more preferable as can be seen from Table \ref{c_i fixed:BF} (the Bayes factor in favour of the Full Gaussian is 142.54 with a non-informative prior for $\alpha$, according to the Importance sampling estimator). This can be characterised as ``very strong'' evidence in favour of the Full Gaussian. The same preference for the Full Gaussian function can be derived by comparing the log-marginal densities for $\alpha \ne 1$ with an informative prior for $\alpha$. 
Looking at the dotted and dot-dashed lines at the left panel of Figure \ref{fig:max_like} we can conclude that Full Gaussian gives a better fit to the data. Also, $P[D_{15,20}>0|\mathbf{D}]=0.84$, $P[D_{15,20}>4.4|\mathbf{D}]=0.66$ and $P[D_{15,20}>7.3|\mathbf{D}]=0.50$. Again, this is ``strong'' evidence in favour of the Full Gaussian. Furthermore, both information criteria favour the Full Gaussian model. Finally, even by eye, the Full Gaussian model produces better results in the second column of Figure \ref{fig1}. Therefore, the fact that we included the $c_i$ as free parameters, did not alter the outcome compared to the analysis in Section \ref{mod_sel_fix} where the $c_i$ are fixed. 

Regarding the thermal and non-thermal heat fluxes, from the right panel of Figure \ref{fig:max_like} it is apparent that there is a slight better fit to the data with the $\alpha>1$ hypothesis. Nevertheless, this improvement in fit does not seem to be substantial. We have also calculated $P[D_{17,18}>0.5|\mathbf{D}]=0.50$, $P[D_{19,17}>0.5|\mathbf{D}]=0.50$ and $P[D_{19,18}>1.0|\mathbf{D}]=0.50$. Otherwise, $P[D_{17,18}>0|\mathbf{D}]=0.52$, $P[D_{19,17}>0|\mathbf{D}]=0.52$ and $P[D_{19,18}>0|\mathbf{D}]=0.55$. According to Table \ref{post_table} none of them is ``worth more than a bare mention''. On the other hand, both information criteria favour $M_{17}$, just as when $c_i$ are fixed. Moreover, the marginal densities of $M_{17}, M_{18}, M_{19}$ are very similar to each other, something that supports the idea that the difference between the models is ``not worth more than a bare mention''. Finally, according to the quantiles of Table \ref{post_interv}, the $95\%$ credible interval with informative prior for $\alpha$ contains 1, whereas the $95\%$ credible interval with non-informative prior for $\alpha$ does not contain 1. In other words, $P(\alpha>1)$ is close to 0.50 with informative prior for $\alpha$, whereas it is close to 1 with non-informative prior for $\alpha$. 
%Using an informative prior for $\alpha$ and the Full Gaussian non-thermal heat flux, it can be concluded from Table \ref{c_i fixed:BF} that the $\alpha>1$ hypothesis is slightly better than the other two. But this evidence is not that strong and it can be characterised as ``not worth more than a bare mention''. For the Half Gaussian model the same picture is obtained: the data we have cannot distinguish which of the thermal or non-thermal heat fluxes is more dominant, if either. 
%Looking at the bottom panel of Figure \ref{fig:max_like} it is apparent that there is a slight better fit to the data with the $\alpha>1$ hypothesis. Nevertheless, this improvement in fit does not seem to be substantial. For example, we have calculated $P[D_{17,18}>0.5|\mathbf{D}]=0.50$, $P[D_{19,17}>0.5|\mathbf{D}]=0.50$ and $P[D_{19,18}>1.0|\mathbf{D}]=0.50$. Otherwise, $P[D_{17,18}>0|\mathbf{D}]=0.52$, $P[D_{19,17}>0|\mathbf{D}]=0.52$ and $P[D_{19,18}>0|\mathbf{D}]=0.55$. According to Table \ref{post_table} none of them is ``worth more than a bare mention''. 
Therefore, we reach the same conclusion as in Section \ref{mod_sel_fix}: although it is clear that the Full Gaussian function in preferable, the data are not sufficient in order to distinguish between the different heating mechanisms.

\subsubsection{Parameter estimation}

All the estimations from the posterior distributions of the parameters can be viewed in Table \ref{c_i fix}, using the Full Gaussian function. We estimate the mean of $L$ to be 31.6 Mm, the mean of the total non-thermal heat flux is $31.99 \times 10^9 $ ergs $\mbox{cm}^{-2}$, the mean time for the peak of this function is $864.6$ s after the beginning of the flare, the mean of the standard deviation of the Full Gaussian is $92.6$ s, the mean of the $\alpha$ parameter is $23.51$, the mean background heating rate is $0.12 \times 10^{-5}$ ergs $\mbox{cm}^{-3}$ $\mbox{s}^{-1}$ and the radius of the loop is $6.7$ Mm. 

Regarding the $c_i$ ratios, the mean of the ratio between the radiative loss rate of the transition and the corona ($c_1$) is $2.07$, the mean of the ratio between the average coronal temperature and the apex temperature ($c_2$) is $0.86$ and the mean of the ratio between the coronal base temperature and the apex temperature ($c_3$) is $0.75$. Once again, we should bear in mind that although the estimation for $\alpha$ is greater than unity ($23.51$) with non-informative prior, a more detailed analysis shown in Section \ref{mod_sel_par} suggests that we cannot distinguish which of the thermal or non-thermal heat fluxes is more dominant. Furthermore, assuming informative prior for $\alpha$, $\alpha$ is greater than unity with probability $57.13\%$. 
%The second column of Figure \ref{fig1} shows the best fit of the parameters regarding the posterior distributions for both the Full and Half Gaussian functions. 

%As in Section \ref{par_est_fix}, we want to check the sensitivity of the analysis regarding the prior information for $\alpha$. Comparing Table \ref{c_i fix} with the respective table we get when we use an informative prior for $\alpha$, then the only difference worth mentioning is the $\mathcal{F}'_{tot}$ parameter which will give a $95\%$ highest posterior density interval $[37.33,318.38]$ for an informative prior for $\alpha$ and $[3.26,119.65]$ for a non-informative prior for $\alpha$.  

\subsection{Comparison Between the Hypotheses: $c_i$ Fixed and $c_i$ Free Parameters.}

\begin{figure}
   \centering
   \includegraphics[scale=0.38]{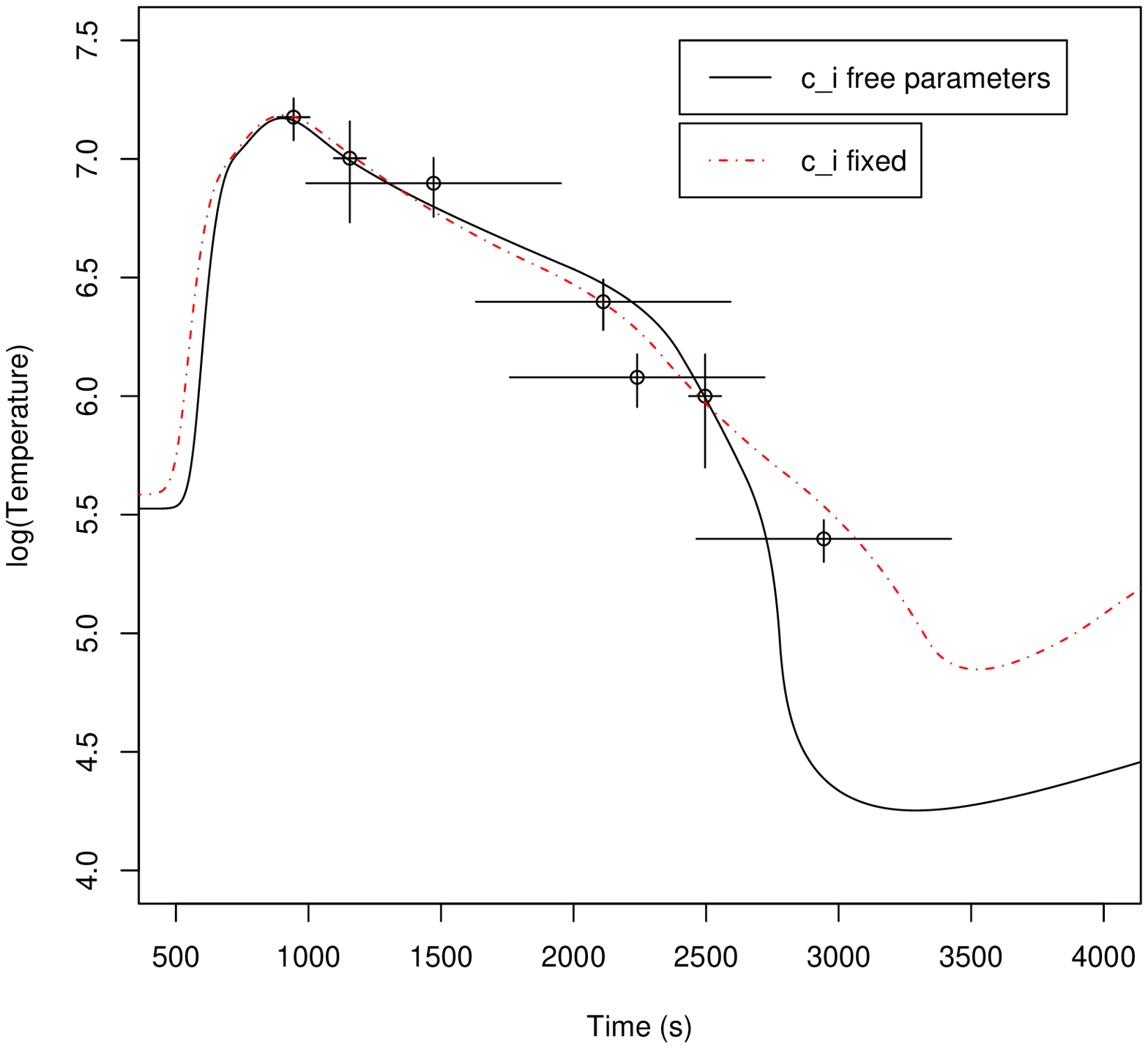} \\
   \includegraphics[scale=0.38]{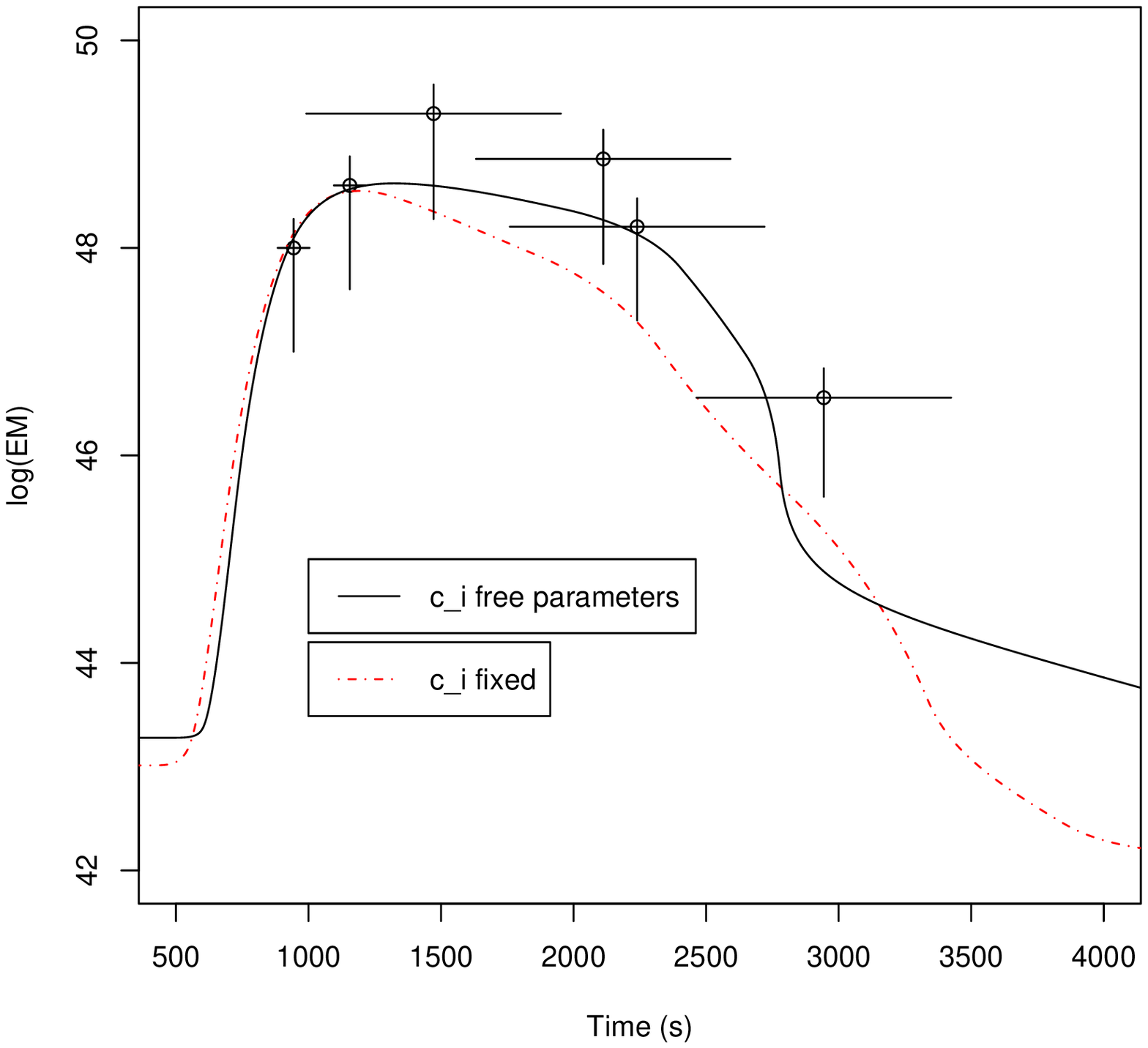} \\
   \includegraphics[scale=0.38]{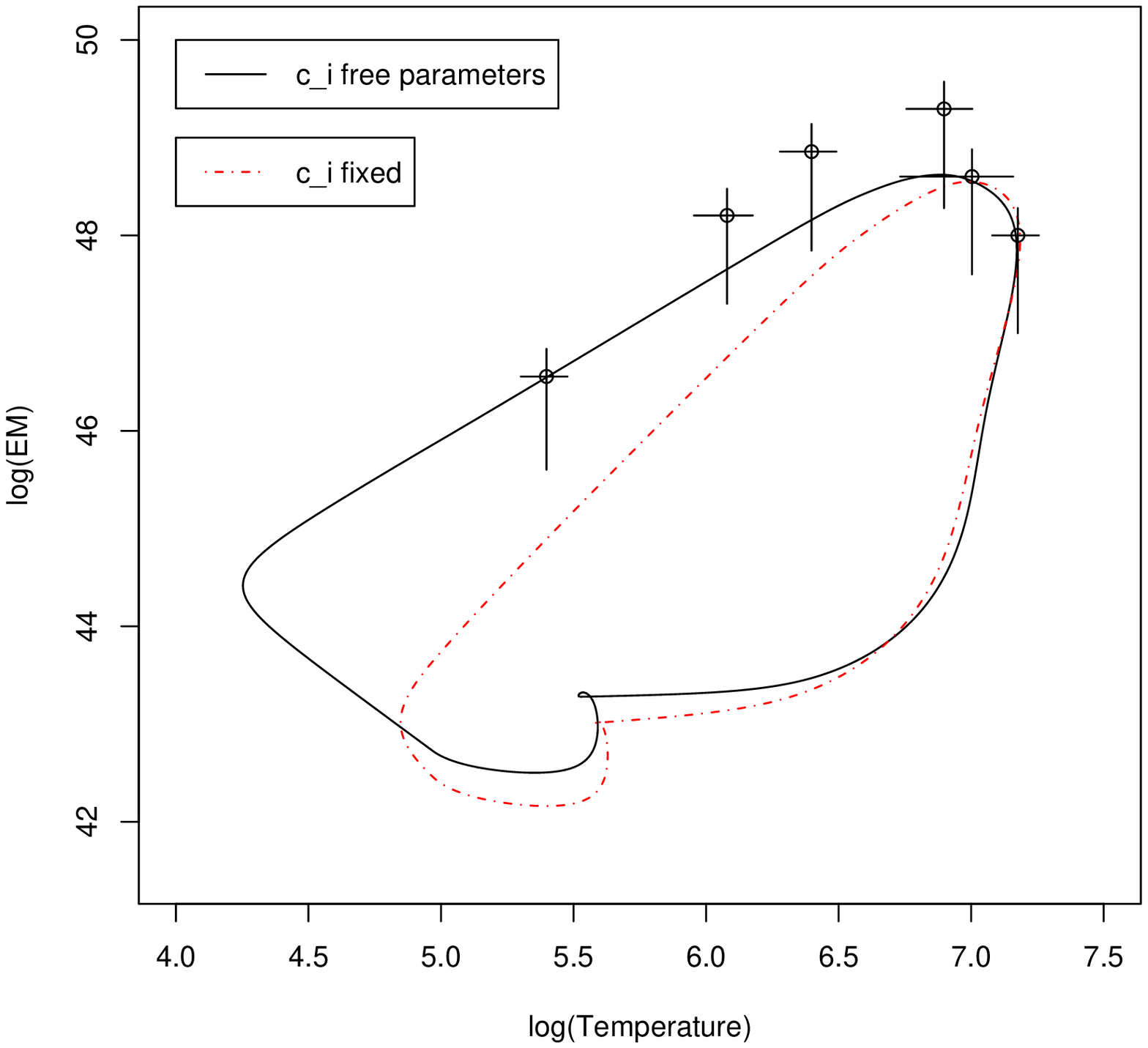}
   \caption{Temperature evolution ({\it first row}), emission measure evolution ({\it second row}) and temperature against emission measure ({\it third row}) using the EBTEL model that best reproduced the observations. Solid lines represent the fit when $c_i$ are floating as free parameters, whereas dot-dashed lines assume $c_i$ to be fixed numbers. Each curve depicts the best fit that maximised the posterior distribution of the Full Gaussian statistical models. A {\bf non-informative} prior for $\alpha$ was employed. Temperature is measured in K, while emission measure is in $\mbox{cm}^{-3}$.}
   \label{te_vs_ti_noninf_fix_param_new}
\end{figure}

Considering the non-thermal heat flux, both hypotheses ($c_i$ fixed and $c_i$ free parameters) propose the Full Gaussian statistical model. However, none of them can distinguish which of the two heating mechanisms is dominant (if any). Regarding Question 3, from Table \ref{c_i fixed:BF} we reach different conclusions depending on the estimation method for the marginal distribution: Importance sampling does not seem to favour any hypothesis, but the other two methods favour the hypothesis $c_i$ free parameters. This can be concluded when a comparison between models $M_{10}$ and $M_{20}$ (or $M_6$ and $M_{16}$) is made. 
On the other hand, the posterior deviances of Figure \ref{fig:max_like} depict preference to the models where $c_i$ are set as free parameters and the form of the non-thermal heat flux is Full Gaussian. We have also calculated $P[D_{10,20}>19.2|\mathbf{D}]=0.50$, $P[D_{10,20}>0|\mathbf{D}]=0.995$ and $P[D_{10,20}>4.4|\mathbf{D}]=0.98$ which indicates a ``very strong'' evidence in favour of model $M_{20}$ against model $M_{10}$. Similar conclusions can be driven with the information criteria, as both of them show clear preference to the $c_i$ free parameter models. Last but not least, the difference between the two hypotheses in Figure \ref{te_vs_ti_noninf_fix_param_new} does not seem to be very important for the thermal evolution (first row), at least by eye. However, for the data-set under consideration, the difference is more profound for the emission measure evolution (second row). This indicates that the $c_i$ parameters affect the emission measure values more than they affect the temperature values. The fact that there are better values for $c_i$ than $4$, $0.87$ and $0.72$ is even more clear in the third row where temperature is plotted against emission measure. 

All the above indicate that the values introduced for $c_i$ when they are fixed are not very good, in terms of maximising the likelihood function. The information added when $c_i$ are free parameters is greater than the price we have to pay for introducing three additional parameters.

\section{Discussion}
\label{discussion}

In order to reveal the mysteries of the Sun, we can break down our investigations into three major stages: (\emph{i}) constructing theoretical models, (\emph{ii}) gathering observations, and (\emph{iii}) applying a statistical analysis in order to compare different statistical models and/or to restrict the parameters of the models. All of them are of equal importance and give us confidence in our results. 

Making inferences solely from the mean or mode of Table \ref{c_i fix} would warrant utmost criticism. For model comparison purposes the $95\%$ credible interval is more robust. However, even with the $95\%$ credible interval we can not compare with the $\alpha=1$ hypothesis. Also, we do not directly include the likelihood function in our calculations. 
Model comparison techniques using Bayesian statistics (\emph{e.g.} Bayes factor) take into account both the spread of the parameter posterior distributions and the information of the likelihood function. However, Bayes factor is sensitive to the choice of prior distributions and great attention should be paid when applying this method. On the other hand, techniques that are based only on posterior likelihood distributions do not depend on the prior distributions of the parameters --- at least not directly. We strongly encourage researchers to employ and compare various statistical techniques when embarking on the subtle topic of model selection.

In this paper, temperature and emission measure profiles produced by the EBTEL model were compared with solar flare observations. The data distribution, the parameter set and the priors employed in this analysis were described in detail. The form of thermal and non-thermal heat input is much better described using Full Gaussian energy profile than Half Gaussian, which is what \citet{raftery08} also used. 

Apart from choosing which energy profile function was more appropriate for the data-set we analysed, we were also interested in determining which of the thermal or non-thermal heat fluxes was more dominant. The data obtained were not able to provide an answer with great confidence. More data-points may be required in order to address this question. It has also been suggested that it might be possible to have mostly non-thermal heating in the impulsive phase and mostly thermal heating thereafter. To test this, Equation (\ref{alpha_param_hest_rate}) can be replaced by: 
\begin{displaymath}
Q(t) = \left\{ 
\begin{array}{ll}
\alpha_1 \mathcal{F}_1(t) + B, \quad t \le t_1 \\
\alpha_2 \mathcal{F}_1(t) + B, \quad t > t_1,
\end{array}
\right.
\end{displaymath}
where $\alpha_1, \alpha_2$ and $t_1$ parameters under consideration. If the above statement were true then one would expect that $t_1$ should be close to where the temperature peak of Figure \ref{fig1} is observed, as well as $\alpha_1<1$ and $\alpha_2>1$.

The $c_i$ parameters were given fixed values in the \citet{kpc08} paper so that the 0D HD model will approximate the 1D HD model. If we assume that the range of these parameters provides sufficient approximations, then we should include them as free parameters. In any case, we disagree with fixing some parameters to certain values, in order to reduce the parameter set, as this might affect the results --- unless we have high confidence about these fixed numbers. 

The fact that the Bayes factor is not so decisive in choosing between the hypotheses $c_i$ fixed or free parameters is partially an outcome of the conservative prior distributions we have chosen for the $c_i$ parameters (see Section \ref{priors}). If we had added more information in the prior distribution of $c_2$, say $\mathcal{U}(0.5,1)$ or even better the Beta distribution $\mathcal{B}(38.49,5.75)$, this would have been in favour of the $c_i$ parameters hypothesis. In comparison, posterior likelihood techniques and information criteria clearly show that $c_i$ should be set as free parameters. 
More improved estimations for these parameters can be seen from the mean or mode of Table \ref{c_i fix} (for the particular data-set we analyse).

Finally, an obstacle presented in this analysis was that of constraining the profiles produced by the EBTEL model. For example, we might not want to restrict the initial values of temperature, density and/or emission measure profiles. This can produce model profiles that are closer to the data profiles, but the initial values might take exceptionally high numbers. For instance, we had undertaken the same analysis without fixing the initial values of temperature, density and emission measure. This resulted in model profiles with initial temperatures\footnote{Initial temperature is assumed to be at the base of the transition region.} of $\sim 3$~MK, initial electron density of $\sim 500 \times 10^7$~$\mbox{cm}^{-3}$ and initial emission measures of $\sim 3 \times 10^{47}$~$\mbox{cm}^{-3}$. This was because the background heating rate was three orders of magnitude higher than that in Table \ref{c_i fix}. Apart from unrealistic estimations of the parameters of interest, this could have also led to unreliable Bayes factor estimations with false conclusions regarding the three posed questions in the beginning of Section~\ref{results_raftery}. Naturally, the quality of the output of the analysis is dependent of the quality of the input. 

Apart from improved statistical techniques, of equal importance is that improvement upon the observations should be made. This means that future missions with new instrumentation should provide data-sets with a large enough number of observations in order to distinguish between different heating mechanisms. The data-set under consideration provided information only upon the decay phase of the temporal evolution. However, a better time resolution for the rise phase of the temperature will be needed in order to provide a better estimate for the form of thermal/non-thermal heat flux. And of course, a large sample of solar flares will be required. 

An assumption made in this paper is that the thermal and non-thermal fluxes have the same form, based on a lack of  information on the thermal distribution. However, it would be interesting to test fluxes of different forms that do not depend on each other. Additionally, a further improvement in the EBTEL model is required regarding the non-thermal heat flux, as it is efficient for gentle chromospheric evaporation but suffers from inadequately representing explosive chromospheric evaporation. Last but not least, several other forms of heating input, like proton beams, could be included in an attempt to make the model more realistic.

\acknowledgments
SA has been supported by a STFC grant. CLR is supported by an ESA/Prodex grant, administered by Enterprise Ireland. The authors would like to express their gratitude to the referee for the useful comments.

%\bibliographystyle{spr-mp-sola}
%\bibliography{biblio}

\bibliographystyle{aa}
\bibliography{biblio}{}

\end{document}